\documentclass[twocolumn]{aastex62} 
\usepackage{xcolor, bm, amsmath, enumitem}
\usepackage[T1]{fontenc} 
\usepackage{array,multirow} 
\usepackage[hang]{footmisc}
\usepackage{mfirstuc}
\usepackage{array}
\newcolumntype{H}{>{\setbox0=\hbox\bgroup}c<{\egroup}@{}}

\newcommand{\rutgers}{Rutgers University, Department of Physics and Astronomy, 136 Frelinghuysen Road, Piscataway, NJ 08854, USA}
\newcommand{\utaustin}{Department of Astronomy, The University of Texas at Austin, 2515 Speedway, Stop C1400, Austin, TX 78712-1205, USA}

\newcommand{\msun}{$M_\odot$}
\newcommand{\mdot}{$\dot{M}_\star$}
\newcommand{\vsini}{$v \,\mathrm{sin}(i)$}

\newcommand{\vsys}{$v_\mathrm{sys}$}

\newcommand{\hii}{H\,\textsc{ii}}
\newcommand{\hi}{H\,\textsc{i}}
\newcommand{\hei}{He\,\textsc{i}}
\newcommand{\heii}{He\,\textsc{ii}}
\newcommand{\teff}{$T_\mathrm{eff}$}
\newcommand{\logg}{$\log(g)$}

\newcommand{\lstar}{$L_\star$}
\newcommand{\mstar}{$M_\star$}

\newcommand{\zsun}{$Z_\odot$}

\newcommand{\ebv}{$E(B-V)$}

\newcommand{\fesc}{$f_\mathrm{esc}$}
\newcommand{\hbeta}{H$\beta$}
\newcommand{\oiii}{[O\,\textsc{iii}]}
\newcommand{\oiiiwave}[1]{[O\,\textsc{iii}]$\,\lambda$#1}
\newcommand{\oiiidoub}[1]{[O\,\textsc{iii}]$\,\lambda\lambda$#1}
\newcommand{\otemp}{T(O$^{++}$)}
\newcommand{\qh}{$Q$(H)}
\newcommand{\qhei}{$Q$(He)}

\newcommand{\kms}{km\,s$^{-1}$}

\newcommand{\tlusty}{\textsc{tlusty}}
\newcommand{\cloudy}{\textsc{Cloudy}}
\newcommand{\pycloudy}{\textsc{PyCloudy}}
\newcommand{\pyneb}{\textsc{PyNeb}}

\shorttitle{The Ionizing Spectrum of an Extremely Metal-Poor O Star}
\shortauthors{Telford et al.}

\begin{document}

\title{The Ionizing Spectra of Extremely Metal-Poor O Stars: Constraints from the Only \hii{} Region in Leo~P}

\author[0000-0003-4122-7749]{O. Grace Telford}
\affiliation{\rutgers}
\email{grace.telford@rutgers.edu}
\author[0000-0001-5538-2614]{Kristen B. W. McQuinn}
\affiliation{\rutgers}
\author[0000-0002-0302-2577]{John Chisholm}
\affiliation{\utaustin}
\author[0000-0002-4153-053X]{Danielle A. Berg}
\affiliation{\utaustin}


\begin{abstract}

Metal-poor, star-forming dwarf galaxies produce extreme nebular emission and likely played a major role in cosmic reionization. 
Yet, determining their contribution to the high-redshift ionizing photon budget is hampered by the lack of observations constraining the ionizing spectra of individual massive stars more metal-poor than the Magellanic Clouds (20$-$50\%\,\zsun{}).
We present new Keck Cosmic Web Imager (KCWI) optical integral field unit spectroscopy of the only \hii{} region in Leo P (3\%\,\zsun{}), which is powered by a single O star.
We calculate the required production rate of photons capable of ionizing H and He from the observed H$\beta$ and He\,\textsc{i}\,$\lambda$4471 emission-line fluxes.
Remarkably, we find that the ionizing photon production rate and spectral hardness predicted by a \tlusty{} model fit to the stellar SED agrees with our observational measurements within the uncertainties.
We then fit \cloudy{} photoionization models to the full suite of optical emission lines in the KCWI data and show that the shape of the same \tlusty{} ionizing continuum simultaneously matches lines across a wide range of ionization energies.
Finally, we detect O\,\textsc{iii}]  and N\,\textsc{iii}] nebular emission in the Hubble Space Telescope far-ultraviolet spectrum of the Leo~P \hii{} region, and highlight that the rarely observed N\,\textsc{iii}] emission cannot be explained by our \cloudy{} models. 
These results provide the first observational evidence that widely used, yet purely theoretical, model spectra accurately predict the ionizing photon production rate from late-O stars at very low metallicity, validating their use to model metal-poor galaxies both locally and at high redshift.

\end{abstract}


\section{Introduction\label{sec:intro}}

\subsection{Ionizing Photon Production by Metal-Poor Massive Stars during the Epoch of Reionization}

Galaxies in the early universe were more metal-poor than their present-day counterparts, as few stellar generations had time to deposit newly formed metals into their interstellar media (ISM).
Metal-poor massive stars were therefore important sources of feedback that regulated galaxy assembly at high redshift ($z$).
Ionizing photons escaping those galaxies via channels carved in the dense ISM by stellar feedback  \citep[e.g.,][]{wise09, erb15} are widely thought to have reionized the intergalactic medium (IGM) by $z \sim 6$ \citep{becker01}.
While it is likely that the lowest-mass, most metal-poor galaxies produced most of the Lyman continuum (LyC) photons at very high redshift \citep[e.g.,][]{dayal18, chisholm22}, the importance and timing of contributions from more massive galaxies and quasars to cosmic reionization is still debated from both theoretical and observational perspectives \citep[e.g.,][]{ouchi09, robertson15, finkelstein19, naidu20}.

Recent observations of high-$z$ galaxies from the new James Webb Space Telescope support the idea that very metal-poor massive stars in blue, faint dwarf galaxies played a large, perhaps dominant, role in reionization \citet{endsley22, nanayakkara22, topping22}.
However, inferring the ionizing photon production by massive stars in these metal-poor, early galaxies requires the use of stellar population synthesis (SPS) models \citep[e.g.,][]{tinsley80, conroy13, chisholm19} to translate the observations into physical quantities. 
These codes are extremely powerful, but ultimately are limited by the quality of the stellar evolution models and spectral templates that they adopt. 
Stellar model uncertainties propagate to the accounting of galaxies' contributions to the high-$z$ ionizing photon budget as a function of their mass.

At present, stellar models in the very low-metallicity ($Z$) regime have not been observationally validated.
The predictions of these models are typically calibrated against large samples of massive stars in Local Group galaxies \citep[e.g.,][]{massey13}.
Individual stars in the Milky Way and in the lower-$Z$ Large and Small Magellanic Clouds (LMC and SMC; 20\%\,\zsun{} and 50\%\,\zsun{}, respectively; \citealt{dufour84}) are close (and therefore bright) enough to routinely observe spectroscopically, which is essential to determine key stellar and wind properties. 
But at lower $Z$, massive OB stars are only found in star-forming dwarf galaxies at much larger distances ($\gtrsim$\,1\,Mpc).
Thus, spectroscopy of individual metal-poor O stars is expensive and rare, and theoretical stellar models have not been directly tested against observations below the 20\%\,\zsun{} of the SMC.

Empirically constraining the ionizing spectra of massive stars is a particularly daunting challenge.
Stellar atmosphere models fit to the observed spectra and photometry of individual stars \citep[e.g.,][]{simon-diaz20} can be extrapolated blueward of 912\,\AA{} to predict the ionizing spectrum of a star, given its observed properties at longer wavelengths. 
But in the local universe, direct observation of LyC radiation is impossible due to strong absorption by \hi{} in the Milky Way.
At moderate $z$, LyC escaping entire galaxies can be observed (though these measurements are difficult; e.g., \citealt{pahl21, flury22}), but this does not constrain the emergent ionizing flux from individual stars.
Nebular emission from \hii{} regions powered by single stars has long been recognized as a useful diagnostic of the shape (hardness) of the star's ionizing spectrum, which is tightly related to its effective temperature (\teff{}; e.g., \citealt{kennicutt00, oey00, dors03, zastrow13}).
However, this technique has only been applied to single-star \hii{} regions in the Milky Way and Magellanic Clouds -- never at lower $Z$.

\subsection{Leo~P and its Only Known O Star}

\begin{figure*}[!htp]
\begin{centering}
  \includegraphics[width=0.6\linewidth]{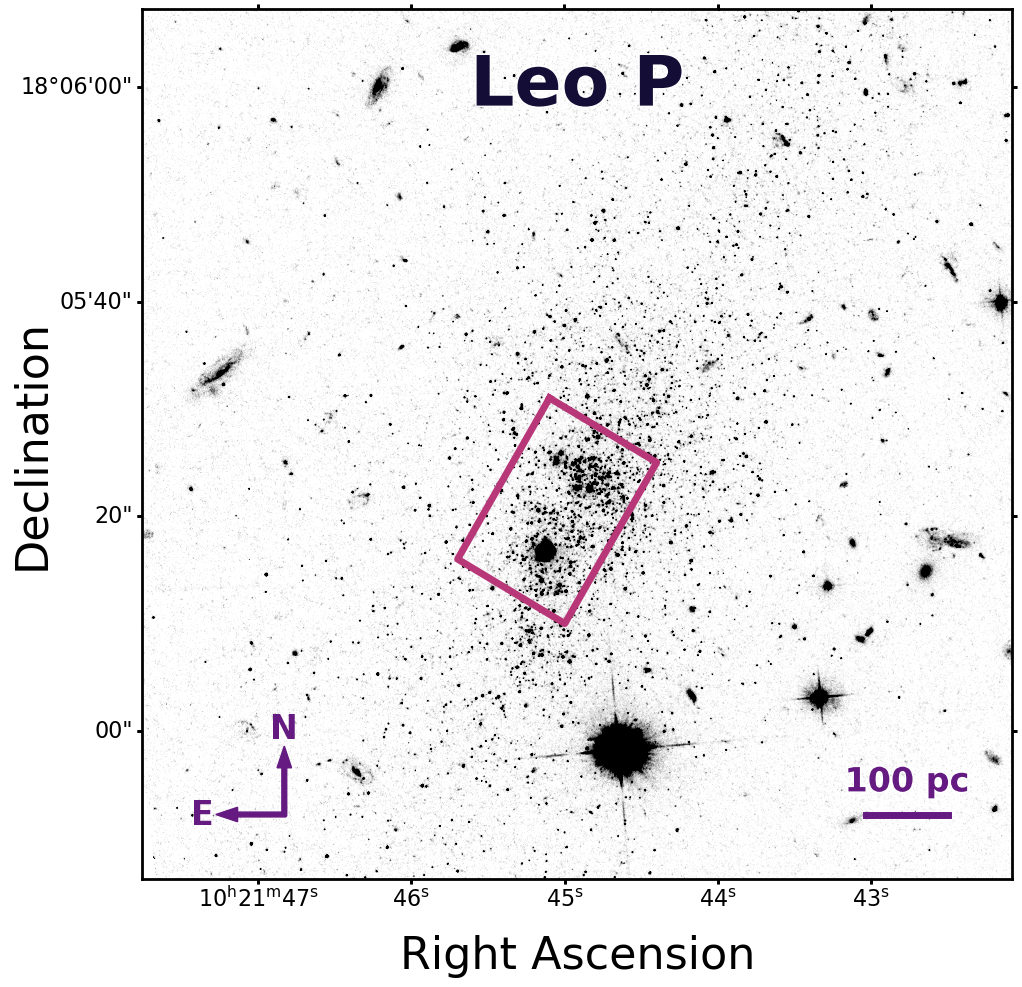}
\caption{\textbf{The Keck/KCWI field of view shown on a HST image of Leo~P.} 
A mosaic in the F475W filter approximately 1\farcm5 on a side is shown in greyscale, with the pink rectangle enclosing the field of view covered by the Keck/KCWI observations shown in Figure~\ref{fig:kcwi_cubes} below. 
The KCWI field was chosen to cover the only \hii{} region in Leo~P and as many of its young stars observed to lie near the upper main sequence in the HST color-magnitude diagram \citep{mcquinn15f} as possible. 
\label{fig:leop_hst}}
\end{centering}
\end{figure*}

\begin{table}
\tabcolsep=0.5cm
\begin{center}
\caption{Properties of Leo~P and LP26.\label{tab:properties}} 
\begin{tabular}{llc}
\hline
\multicolumn{3}{c}{\textbf{Leo~P}} \\
R.A. (J2000) & 10:21:45.0 & (1) \\
Decl. (J2000)  & +18:05:01 & (1) \\
12+$\log$(O/H) & $7.17\pm0.04$ & (2) \\
\vsys{} (km\,s$^{-1}$) & $260.8\pm2.5$ & (3) \\
Distance (Mpc) & $1.62\pm0.15$ & (4) \\
\mstar{} (\msun{}) & 5.6$^{+0.4}_{-1.9}$ $\times$ 10$^5$ & (4) \\
$M_\mathrm{H\,\textsc{i}}$ (\msun{}) & 8.1 $\times$ 10$^5$ & (4) \\
SFR$_{\mathrm{H}\alpha}$ (\msun{}\,yr$^{-1}$) & 4.3 $\times$ 10$^{-5}$ & (4), (5) \\
\ebv{} (mag) & 0.014 & (6) \\
\multicolumn{3}{c}{\textbf{LP26}} \\
R.A. (J2000) & 10:21:45.12 & (7) \\
Decl. (J2000) & +18:05:16.93 & (7) \\
Spectral Type & O7$-$8~V & (4) \\
\teff{} (kK) & 37.5$^{+5.9}_{-5.5}$ & (7) \\
$\log(g/\mathrm{cm\,s}^{-2})$ & 4.0$^{+0.19}_{-0.24}$ & (7) \\ 
$\log(L_\star/L_\odot)$ & $5.1 \pm 0.2$ & (7) \\
\vsini{} (km\,s$^{-1}$) & $370 \pm 90$ & (7) \\
\end{tabular}
\end{center}
\tablecomments{\hi{} emission centroid coordinates are given for Leo~P. Right ascension is reported in hours, minutes, seconds, and declination is in degrees, arcminutes, arcseconds. \ebv{} is due to the Milky Way foreground, and does not include dust in Leo~P. References: (1) -- \citet{giovanelli13}; (2) -- \citet{skillman13}; (3) -- \citet{bernstein-cooper14}; (4) -- \citet{mcquinn15f}; (5) -- \citet{green15}; (6) -- \citet{rhode13}; (7) -- \citet{telford21}.}
\end{table}

Leo~P was detected as an \hi{} source \citep{giovanelli13} in the ALFALFA survey \citep{giovanelli05, haynes11}.
Followup ground-based imaging by \citet{rhode13} confirmed a stellar counterpart and an \hii{} region in the galaxy, and suggested that it is located outside the Local Group.
\citet{skillman13} then obtained optical spectroscopy of the \hii{} region and determined ${12+\log{(\mathrm{O/H})} = 7.17 \pm 0.04}$, or 3\%\,\zsun{} (relative to the solar oxygen abundance of 8.69; \citealt{asplund09}), making Leo~P an extremely metal-poor (XMP) galaxy (defined as ${12+\log{(\mathrm{O/H})} \leq 7.35}$ or $\lesssim 5\%\,$\zsun{}; e.g., \citealt[][]{mcquinn20}).

As the most metal-poor, star-forming XMP galaxy close enough that its stars can be resolved (Table~\ref{tab:properties}), Leo~P has been of great interest as a benchmark for the astrophysics of very metal-poor O stars.
Stellar photometry from Hubble Space Telescope (HST) imaging of Leo~P presented by \citet{mcquinn15f} enabled a census of stars near the \hii{} region, which revealed only one source on the main sequence bright enough to be an O star producing substantial ionizing flux. 
\citet{evans19} used VLT/MUSE optical integral field unit (IFU) spectroscopy to search for additional massive stars in Leo~P.
These observations confirmed the O type of the lone star embedded in the \hii{} region, dubbed LP26, via detection of photospheric \heii{} absorption. 
However, the signal-to-noise ratio (SNR) and low resolution ($R\sim2000$) of the MUSE data combined with heavy contamination of the \hei{} photospheric absorption by nebular emission prevented a detailed analysis of the stellar properties.
No evidence of any other stars earlier than spectral type B was found, so LP26 remains the only known O star in Leo~P to date. 

Recently, \citet{telford21} presented far-ultraviolet (FUV) spectroscopy of LP26 obtained with the Cosmic Origins Spectrograph (COS) on HST.
Analysis of the photospheric and wind line profiles revealed that the star is a fast rotator, with an estimated \vsini{} of 370\,\kms{}, and that it is driving an optically thin stellar wind, requiring a very low mass-loss rate (\mdot{}). 
Both the FUV photospheric line strengths and the shape of the star's observed spectral energy distribution (SED) were found to be consistent with a $T_\mathrm{eff}\sim 37.5$\,kK main-sequence star, in line with expectations for its estimated O7$-$8~V spectral type \citep[from the absolute F475W magnitude;][]{mcquinn15f}.
Thus, LP26 is a unique example of a well-characterized XMP O star, and the nebular emission from the \hii{} region that it powers is a powerful diagnostic of the star's ionizing spectrum and H-ionizing photon production rate, \qh{}.

In this paper, we present new optical IFU observations of the \hii{} region in Leo~P obtained with the Keck Cosmic Web Imager (KCWI; \citealt{morrissey18}) on the 10-m Keck~II telescope.
KCWI is optimized for observing in the blue, so these new data cover key stellar photospheric absorption and nebular emission features that are not sampled by the existing MUSE IFU data.
The IFU observations enable a measurement of the \textit{total} flux of each emission line integrated over the entire area of the \hii{} region, avoiding the problem of slit losses.
We compare these observed line fluxes to the nebular emission expected for the ionizing spectra predicted by theoretical stellar atmosphere models matched to the known properties on LP26 (Table~\ref{tab:properties}).
This first test of extremely low-$Z$ spectral models is an important benchmark for analysis of ionizing photon production in metal-poor galaxies, both in the local universe and during the epoch of reionization. 

This paper is organized as follows. 
Section~\ref{sec:data} presents our Keck/KCWI optical and HST/COS FUV observations of the O star and \hii{} region in Leo~P.
Section~\ref{sec:emlines} describes our methods for combining KCWI \hii{} region spectra across different observing runs and for measuring the nebular emission line fluxes and key properties of the \hii{} region from those fluxes. 
Section~\ref{sec:results} then presents our measurements of LP26's ionizing photon production rate and spectral shape.
Section~\ref{sec:discussion} discusses the implications of our findings and highlights unusual aspects of the observed FUV nebular emission.
Finally, our conclusions are summarized in Section~\ref{sec:conclusions}.


\section{Observations and Data Reduction\label{sec:data}}

\subsection{Keck/KCWI Optical IFU Spectroscopy\label{sec:kcwi_data}}

\begin{figure*}[!htp]
\begin{centering}
  \includegraphics[width=0.8\linewidth]{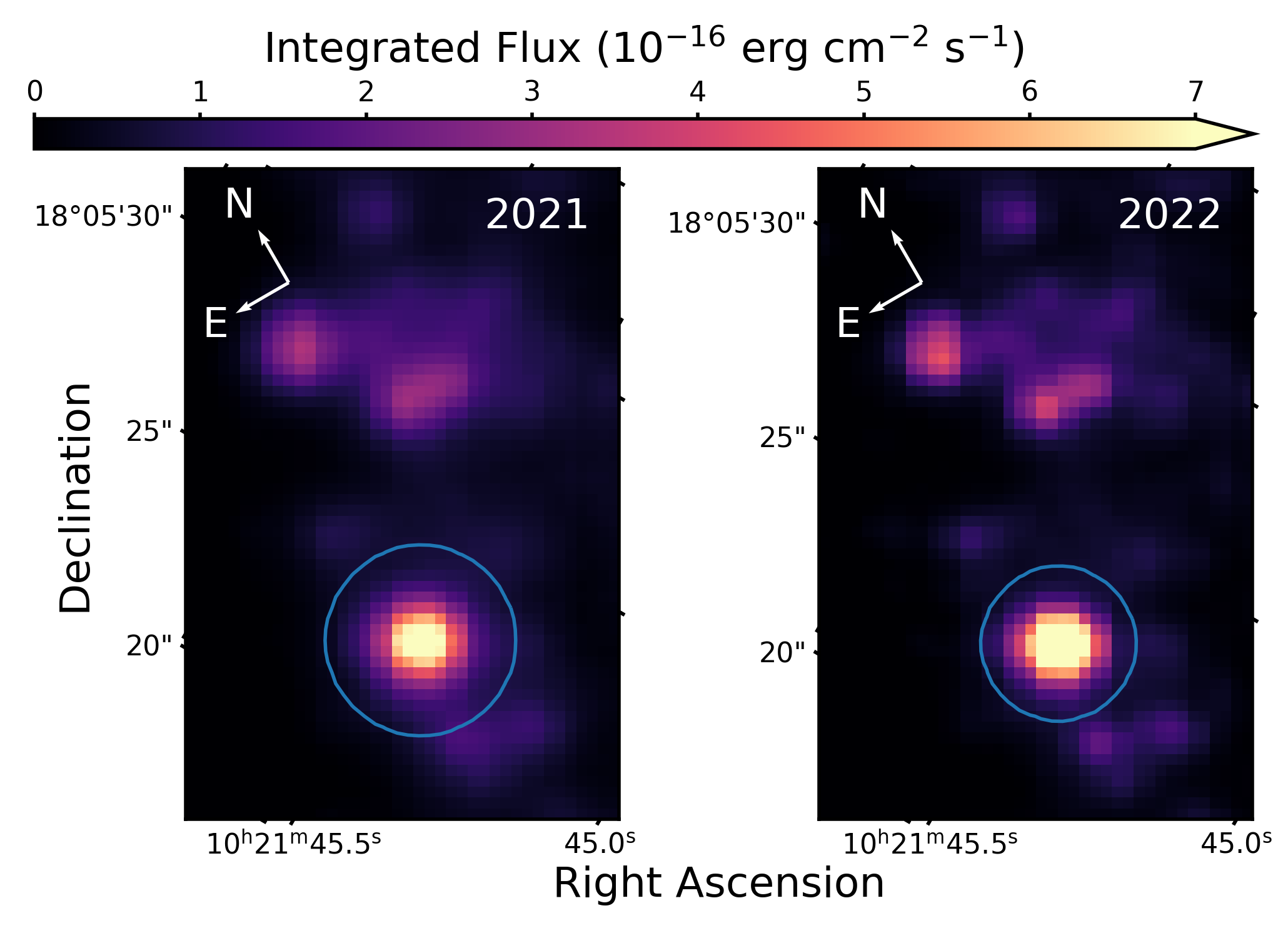}
\caption{\textbf{The coadded KCWI data cubes from the 2021 and 2022 observing runs.} The left and right panels show white-light images constructed by integrating the flux in each spaxel of the coadded data cubes over the full wavelength range covered by our 2021 and 2022 observations, respectively. The same field in Leo~P was observed at different cenwaves and under different seeing conditions during the two observing runs (Section~\ref{sec:kcwi_data}). We therefore do not attempt to combine the data cubes from the two observing runs and instead extract the 1-D \hii{} region spectrum from each separately. Blue circles show the region within which spaxels were summed to construct the total \hii{} region spectrum from each observation (shown in Figure~\ref{fig:extracted_spectra}). The radii are equal to three times the best-fit $\sigma$ of a 2-D Gaussian model fit to the spatial distribution of the H$\beta$ emission, chosen to enclose 99\% of the H$\beta$ flux (Section~\ref{sec:optical_fluxes}).
\label{fig:kcwi_cubes}}
\end{centering}
\end{figure*}

We obtained Keck/KCWI observations of the \hii{} region in Leo~P on January 13, 2021 (Program ID N194; PI: Chisholm) and on January 4, 2022 (N194; PI: Telford).
We used the Medium slicer and BM grating, achieving a resolution of $\sim$\,4000 (75\,km\,s$^{-1}$, or 1.0--1.3\,\AA{} over the wavelength range of our spectra). 
We binned by 2 pixels in both the spatial and spectral dimensions, giving spaxels that are 0\farcs29\,$\times$\,0\farcs68.
We used a central wavelength of 4500\,\AA{} for the 2021 observing run, resulting in wavelength coverage from $\sim$\,4060--4930\,\AA{}, then a central wavelength of 4700\,\AA{} for the 2022 observations to cover the \oiiidoub{4959,\,5007} emission lines.

Figure~\ref{fig:leop_hst} shows the KCWI field of view on a HST image of Leo~P.
The $16\farcs5 \times 20\farcs0$ field of view was oriented at a position angle (PA) of $150^\circ$ to cover the \hii{} region and most of the young stars in the galaxy.
We interleaved observations at the position shown in Figure~\ref{fig:leop_hst} with pointings offset  by 0\farcs7--5\farcs4 to the northeast (moving perpendicular to the $20''$ side of the field of view).
These offset fields sample regions free of bright stars or nebular emission, which we use for background subtraction, while keeping the target \hii{} region in the field of view at all times to maximize the signal-to-noise ratio (SNR).
We required a continuum SNR of $\sim$\,30 for these observations to determine the properties of the central O star powering the \hii{} region from modeling the weak stellar absorption features; that analysis will be presented in a separate paper (Telford et al.\ 2023, in preparation). 
This continuum SNR ensures that even some of the weakest nebular emission lines are detected at the 2-$\sigma$ level (see Table~\ref{tab:obs_lines} and Section~\ref{sec:optical_fluxes} below).

We obtained a total of ten 20-minute exposures during the 2021 observations. Seeing was typically  1\farcs4--1\farcs5 for the first seven exposures, as reported by the DIMM at the telescope. 
However, the seeing became poorer and unstable throughout the final three exposures, so we excluded those frames from this analysis because the \hii{} region flux was spread over a much larger area on the detector and could not be straightforwardly coadded with the higher-quality exposures. 
Conditions were significantly better for the 2022 observations, with seeing stable around 1\farcs2. 
We obtained three 10-minute exposures at the same PA ($150^\circ$) but different 4700\,\AA{} cenwave.
A shorter total exposure time was adequate because our goal was to measure very bright nebular emission lines, not weak stellar continuum absorption features.

For both the 2021 and 2022 observations, we used the standard star Feige~34 to flux-calibrate our spectra.
We observed the flux standard immediately before the first Leo~P science exposure in 2021, and immediately after the last science exposure in 2022. 
Another standard star was observed at the end of the half night in 2021, but is not used in this work because the deteriorating seeing conditions drove us to exclude the Leo~P science exposures closer in time to that standard-star observation.

We reduced the observations using the IDL-based KCWI data reduction ripeline\footnote{\url{https://github.com/Keck-DataReductionPipelines/KcwiDRP}} (DRP; version 1.2.1). 
This pipeline is described in \citet{morrissey18}, but we summarize the key steps here.
Bias and dark frames were subtracted from the raw science images and cosmic rays were rejected.
Continuum bars and arc lamp calibration images were then used to obtain geometry and wavelength solutions, which are required to transform the raw 2-D images into wavelength-calibrated, 3-D data cubes. 
Internal and dome flats were stacked and used to correct the illumination of the science images. 
We used the off-galaxy regions sampled by our offset frames to model the sky background.
All slices that contained bright nebular emission were excluded from the construction of the sky model, and we verified that the resultant sky model image did not contain any unphysical features at the wavelengths of bright emission lines. 
After the sky background was subtracted, a 3-D data cube was constructed and a correction for differential atmospheric refraction was calculated as a function of airmass, position, and wavelength.
Finally, the pipeline provides an interactive tool to fit the inverse sensitivity curve, constructed by comparing a standard star observation to reference spectra from the CALSPEC database\footnote{\url{http://www.stsci.edu/hst/observatory/crds/calspec.html}}. 
We modeled the observed spectra of the standard Feige~34, taking care to mask absorption features, to flux-calibrate our observations of Leo~P.

To coadd the individual exposures, we used CWITools \citep{osullivan20}, a Python package designed to analyze data products from the KCWI and Palomar Cosmic Web Imager data reduction pipelines.
Individual frames were trimmed to exclude the unusable edges of the field of view and wavelength range.
The spatial profile of the \hii{} region in the white light image was modeled as a Gaussian distribution to determine the pixel location of the center, then the known coordinates of the \hii{} region from HST imaging were used to determine a WCS solution for each frame.
The aligned frames were then resampled onto a uniform wavelength and spatial grid, where the output spatial grid pixels are square, as opposed to the native rectangular pixels in the data cubes produced by the KCWI DRP. 
The spatial pixels are 0\farcs29 on a side (corresponding to 2.3\,pc at the distance of Leo~P), and the wavelength sampling is 0.5\,\AA{}.
Finally, the individual frames were averaged, weighted by the exposure time, to produce the final coadd.
The same procedure was applied to the variance and mask files associated with each exposure.
We produced two separate coadds, one for each the 2021 and 2022 observing run, because the conditions were different enough during the two nights that coadding all of the data would be nontrivial and could potentially introduce systematics into our analysis. 
We prefer to construct 1-D \hii{} region spectra from each coadded data cube, then combine the 1-D spectra, rather than attempt to coadd the two data cubes with different spatial resolution. 

Figure~\ref{fig:kcwi_cubes} shows white-light images (i.e., the flux integrated over the full wavelength range) produced from the coadded KCWI data cubes from the 2021 (left) and 2022 (right) observing runs. 
The cubes are oriented at the same PA and both cover the field of view indicated by the black box in Figure~\ref{fig:leop_hst}. 
The \hii{} region is the brightest, yellow feature in each image, encircled in blue.
Fainter features in the white-light images are bright stars in Leo~P, all less massive and later-type than the O star, LP26, at the center of the \hii{} region.
The better seeing during our 2022 observing run is evident, as the individual stars are more sharply defined in the right image. 
The blue circles show the regions within which the one-dimensional \hii{} region spectra were constructed from these data cubes; we describe our choice of these region sizes in Section~\ref{sec:optical_fluxes} below. 

\subsection{HST/COS FUV Spectroscopy\label{sec:cos_data}}

We also use emission line detections in the HST/COS FUV spectrum of the \hii{} region in Leo~P, which was obtained as part of HST program GO-15967 (PI: J. Chisholm) and presented in \citet{telford21}.
We refer the reader to that paper for a detailed description of the data reduction, but summarize key information here.
The nebular emission lines fall within the spectral region $1660-1750$\,\AA{}, which is covered by our observations taken with the G160M grating (centered at 1600\,\AA) using the 2\farcs5-diameter COS primary science aperture (PSA).
We averaged the reduced, one-dimensional spectra produced by the CalCOS pipeline (version 3.3.9) in the *x1dsum.fits files from the four visits, weighted by the inverse variance, to produce the final co-added spectrum.
The spectral resolution measured from Gaussian fits to intrinsically narrow Milky Way ISM absorption features is 50\,km\,s$^{-1}$, sampled at 12.23\,m\AA{}\,pixel$^{-1}$.
We bin the COS spectrum by 12 pixels to increase the SNR of the data while ensuring at least two samples per resolution element.
The resultant 0.15\,\AA{}\,pixel$^{-1}$ spectral sampling is finer than the 0.5\,\AA{}\,pixel$^{-1}$ of the KCWI data, but similar in velocity sampling ($\sim 30\,\mathrm{km\,s}^{-1}$).


\begin{figure*}[!htp]
\begin{centering}
  \includegraphics[width=\linewidth]{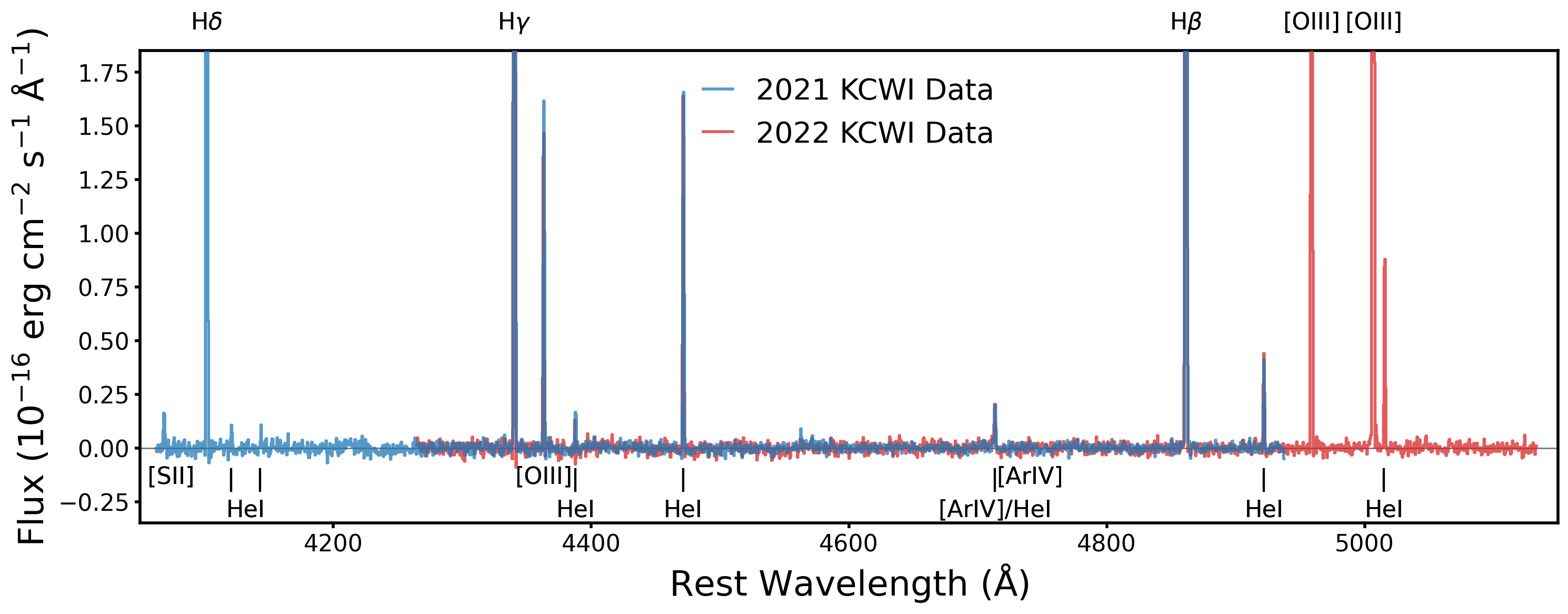}
\caption{\textbf{KCWI optical emission line spectra of the Leo~P \hii{} region.} 
The continuum-subtracted flux is plotted as a function of the rest wavelength (corrected for the velocity measured from the Balmer lines). 
The spectra spanning 4060--4930\,\AA{} (blue line) and 4260--5130\,\AA{} (red line) were observed in 2021 and 2022, respectively.
The strongest emission lines (Balmer lines and \oiiidoub{4549, 5007}) extend beyond the y-axis limit, which was chosen to show the weaker lines detected in the data. 
The continuum SNR is quite high in both spectra, reaching $\sim 15$ for the 2022 observations and $\sim 18$ for the longer-exposure 2021 observations.
Comparing to the horizontal grey line at zero flux demonstrates that our continuum model (particularly for the Balmer absorption features) accurately removes the contribution of the continuum to the emission line fluxes. 
\label{fig:extracted_spectra}}
\end{centering}
\end{figure*}

\section{Emission Line Measurements\label{sec:emlines}}

\subsection{Optical Emission Line Fluxes from KCWI\label{sec:optical_fluxes}}

The KCWI IFU observations offer a key advantage over long-slit spectroscopy: we can measure the total flux in the nebular emission lines across the entire \hii{} region, not just the portion covered by a narrow slit. 
To choose the size of the apertures within which we extract the spaxels that contribute to the 1-D \hii{} region spectra, we first create a narrowband \hbeta{} image from each of the 2021 and 2022 data cubes by summing the flux over the observed wavelength range $4863-4868$\,\AA{} (which covers the \hbeta{} line at the systemic velocity of Leo~P, 261\,km\,s$^{-1}$; Table~\ref{tab:properties}). 
We model the spatial distribution of flux in the H$\beta$ image as a 2-D Gaussian, requiring the standard deviation ($\sigma$) to be equal in both spatial dimensions so that lines tracing a constant flux in the model are circular.
We then sum the spectra in all spaxels falling within $r < 3\sigma$ of the peak of the \hbeta{} flux (blue circles in Figure~\ref{fig:kcwi_cubes}), where $\sigma$ is the best-fit value from the 2-D Gaussian model, to ensure that 99\% of the emission line flux is captured.  
As expected, we find a smaller $\sigma$ for the 2022 observations, which were taken under better seeing conditions: $\sigma_{2021}=2.86$\,pixels (0\farcs83), while $\sigma_{2022}=2.32$\,pixels (0\farcs67). 

For this analysis, we must isolate the nebular line emission from the nebular and stellar continuum flux.
We therefore subtract the continuum level, modeled as a fifth-degree polynomial (with absorption and emission features masked), from the observed one-dimensional \hii{} region spectra.
After this initial removal of the overall continuum shape, we model the stellar Balmer absorption lines (H$\delta$, H$\gamma$, and H$\beta$) with Lorentzian functions, masking the narrow Balmer emission lines superimposed on the broad absorption features.
We then subtract the best-fit Lorentzian profiles to obtain the pure emission spectra of the \hii{} region, with a continuum level of zero and all flux due to the nebular line emission. 
The continuum level is different across in the two \hii{} region spectra from the 2021 and 2022 data cubes because the poorer seeing during the 2021 observations cause more pixels to contribute to the \hii{} region spectrum, which blends with light from nearby stars.

Figure~\ref{fig:extracted_spectra} presents the continuum-subtracted \hii{} region spectra extracted from the coadded 2021 (blue) and 2022 (red) KCWI data cubes, plotted as a function of the rest-frame wavelength (corrected for the velocity measured from the observed wavelengths of the Balmer emission lines).
The strongest emission lines (H$\delta$, H$\gamma$, H$\beta$, and \oiiidoub{4959,\,5007}) extend beyond the upper flux limit in this plot, which was chosen to show the excellent agreement in the weaker emission lines across the two observations.
The horizontal grey line at a flux of 0 is shown for reference, and makes clear that the continuum modeling and subtraction procedure properly removes the continuum contribution to the observed line fluxes, including the stellar Balmer absorption features. 
From the continuum RMS between $4500-4700$\,\AA{}, we estimate SNR of $\sim 15$ for the 2022 observations and $\sim 18$ for the longer-exposure 2021 observations. 

We average the 2021 and 2022 continuum-subtracted, rest-frame, one-dimensional spectra shown in Figure~\ref{fig:extracted_spectra} in the wavelength range where the two datasets overlap, weighting by the inverse variance, to produce the final coadded emission line spectrum that we use throughout the rest of this work. 
We measure the total line fluxes, $F_\lambda$, by simultaneously fitting Gaussian profiles to all observed emission lines.
We allow the velocity of each line to shift by up to half of the 0.5\,\AA{}  pixel width from the global velocity measured from the Balmer lines, corresponding to $\pm 19 \,\mathrm{km\,s}^{-1}$ at the bluest wavelengths.
The observed velocity full width at half maximum (FWHM) of the emission lines has contributions from both the physical velocity dispersion of the emitting gas and the instrumental resolution.
For the Medium slicer and BM grating, the instrumental resolution FWHM of KCWI increases substantially from $R\sim3750$ (80\,km\,s$^{-1}$) at the bluest wavelengths in our data to $R\sim5000$ (60\,km\,s$^{-1}$) at the red end \citep{morrissey18}.
Therefore, we break our spectrum into four regions (split at 4300\,\AA{}, 4600\,\AA{}, and 4900\,\AA{}) and require that all lines within each region have the same velocity FWHM to ensure high-quality fits to the weaker, lower-SNR lines. 
As expected, the best-fit velocity FWHM for the four spectral regions decrease from the bluest to reddest wavelengths: 90.34, 84.68, 77.40, and 66.85\,km\,s$^{-1}$.
The emission line flux is then calculated as the area under the best-fit Gaussian model to each line.

\begin{figure*}[!htp]
\begin{centering}
  \includegraphics[width=\linewidth]{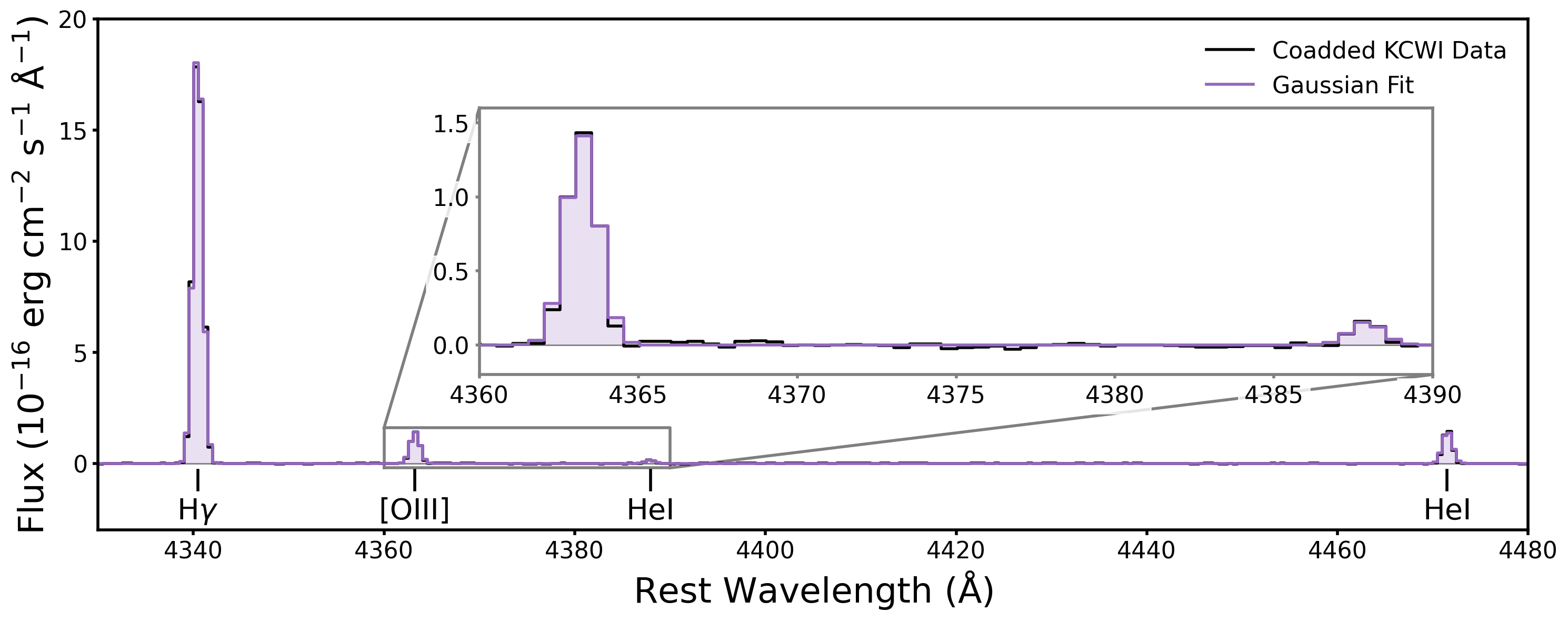}
\caption{\textbf{Example Gaussian fits to nebular emission lines in the coadded KCWI spectrum.} 
The continuum-subtracted emission line flux is plotted as a function of rest-frame wavelength in black, and the best-fit Gaussian emission line model (Section~\ref{sec:optical_fluxes}) is shown in purple. We show only a subset of the lines in the KCWI data to illustrate the fit quality across both strong and weak emission lines. The fits to two of the weaker lines within the grey rectangle are shown in the inset plot.
\label{fig:optical_gaussian_fits}}
\end{centering}
\end{figure*}

Figure~\ref{fig:optical_gaussian_fits} illustrates the quality of our Gaussian fits to the emission lines in the KCWI coadd. 
The observed, continuum-subtracted flux is plotted as a function of the rest-frame wavelength in black, and the best-fit emission line model is plotted in purple.
The velocity FWHM of the emission lines in the spectral region shown were tied together, and this procedure results in excellent fits to both the strongest line (H$\gamma$) and the nearby weaker lines, two of which are shown in greater detail in the inset plot.

The uncertainties on our line flux measurements are calculated by resampling the two flux-calibrated \hii{} region spectra (from the 2021 and 2022 observations) and repeating the normalization, coaddition, and Gaussian fitting procedure 10,000 times.
The resampled flux in each pixel is drawn from a Gaussian distribution centered on the original pixel flux and $\sigma$ equal to the measurement uncertainty.
We find that the uncertainties on the spectra were underestimated relative to the standard deviation in the flux within featureless continuum regions, so choose to scale the uncertainties up by a factor of two to ensure realistic uncertainties on our line flux measurements.
Indeed, we confirm that in those continuum regions, the measured continuum level fell within the 16-84 percentile range of the resampled fluxes in 68\% of pixels after the measurement uncertainties were scaled up.
We adopt half of the difference between the 84 and 16 percentile values of the resampled line flux measurements as the 1-$\sigma$ uncertainties.
Finally, we check that the emission line fluxes and uncertainties measured from the 2021 and 2022 spectra separately agree with the measurements from the coadd for all lines.

\begin{table}
    \caption{Observed Emission Line Fluxes}
    \label{tab:obs_lines}
    \begin{center}
    \tabcolsep=0.25cm
    \begin{tabular}{lccc}
    
     \multicolumn{4}{c}{\textbf{HST/COS FUV}}\\
    Ion & $\lambda_\mathrm{vac}$ & $F_\lambda$ & $I_\lambda$/$I_{\lambda 1666}$ \\
    \hline
O\,\textsc{iii}$]$ & 1660.81& $1.05 \pm 0.22$ & $51.96 \pm 12.44$\\
O\,\textsc{iii}$]$ & 1666.15& $2.02 \pm 0.24$ & $100.00$\\
N\,\textsc{iii}$]$ & 1749& $1.01 \pm 0.33$ & $50.10 \pm 17.59$\\\\
     \multicolumn{4}{c}{\textbf{Keck/KCWI Optical}}\\
    Ion & $\lambda_\mathrm{air}$ & $F_\lambda$ & $I_\lambda$/$I_{\mathrm{H}\beta}$ \\
    \hline 
$[$S\,\textsc{ii}$]$ & 4068.60& $0.23 \pm 0.03$ & $0.44 \pm 0.05$\\
$[$S\,\textsc{ii}$]$ & 4076.35& $0.05 \pm 0.02$ & $0.10 \pm 0.04$\\
H$\delta$ & 4101.73& $14.01 \pm 0.05$ & $26.32 \pm 0.23$\\
He\,\textsc{i} & 4120.82& $0.15 \pm 0.02$ & $0.28 \pm 0.04$\\
He\,\textsc{i} & 4143.76& $0.10 \pm 0.02$ & $0.19 \pm 0.04$\\
H$\gamma$ & 4340.46& $25.29 \pm 0.10$ & $47.42 \pm 0.42$\\
$[$O\,\textsc{iii}$]$ & 4363.21& $1.86 \pm 0.02$ & $3.49 \pm 0.04$\\
He\,\textsc{i} & 4387.93& $0.21 \pm 0.01$ & $0.39 \pm 0.03$\\
He\,\textsc{i} & 4471.49& $1.98 \pm 0.02$ & $3.70 \pm 0.04$\\
$[$Ar\,\textsc{iv}$]$ & 4711.26& $0.05 \pm 0.02$ & $0.10 \pm 0.03$\\
He\,\textsc{i} & 4713.03& $0.24 \pm 0.01$ & $0.46 \pm 0.02$\\
$[$Ar\,\textsc{iv}$]$ & 4740.12& $0.04 \pm 0.01$ & $0.07 \pm 0.02$\\
H$\beta$ & 4861.33& $53.59 \pm 0.43$ & $100.00$\\
He\,\textsc{i} & 4921.93& $0.46 \pm 0.01$ & $0.86 \pm 0.02$\\
$[$O\,\textsc{iii}$]$ & 4958.91& $23.96 \pm 0.06$ & $44.69 \pm 0.38$\\
$[$O\,\textsc{iii}$]$ & 5006.84& $71.95 \pm 0.08$ & $134.12 \pm 1.09$\\
He\,\textsc{i} & 5015.68& $1.08 \pm 0.02$ & $2.02 \pm 0.04$\\
\end{tabular}
    \end{center}
\tablecomments{Vaccuum and air laboratory wavelengths are given for FUV and optical transitions, respectively, all in \AA{}. 
$F_\lambda$ is in units of 10$^{-16}$\,erg\,cm$^{-2}$\,s$^{-1}$. $I_\lambda$ has been corrected for dust extinction and is reported as a percentage of the O\,\textsc{iii}]\,$\lambda$1666 (FUV) or H$\beta$ (optical) intensity.
FUV fluxes are measured within the 1\farcs25-radius COS aperture, while optical fluxes are measured from the coadded 1-D spectra within 3-$\sigma$ apertures as described in Section~\ref{sec:optical_fluxes}. 
The reported N\,\textsc{iii}]\,$\lambda$1749 is the blend of N\,\textsc{iii}]\,$\lambda\lambda$1748.65,\,1749.67.}
\end{table}

Table~\ref{tab:obs_lines} reports the $F_\lambda$ for all emission lines in the coadded KCWI \hii{} region spectrum. 
We also report the ratio of the extinction-corrected intensities ($I_\lambda$, using the total \ebv{} determined from the observed Balmer line ratios in Section~\ref{sec:neb_conditions}) to the \hbeta{} intensity (reported as a \%).
The intensity ratios relative to \hbeta{} are not sensitive to systematic uncertainties in the observed flux calibration or in the ionizing source luminosity, and are therefore used in our comparison to photoionization models in Section~\ref{sec:cloudy} below.

\subsection{FUV Emission Line Fluxes from HST/COS\label{sec:fuv_fluxes}}

\begin{figure*}[!htp]
\begin{centering}
  \includegraphics[width=\linewidth]{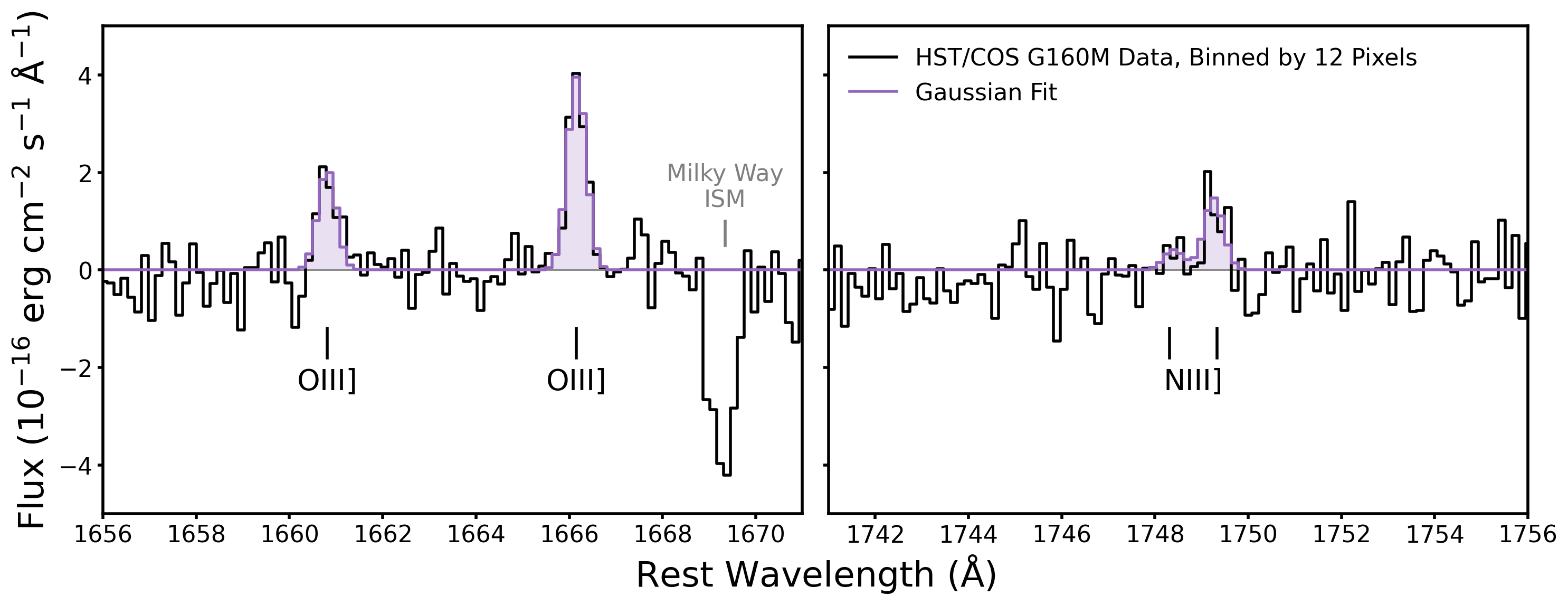}
\caption{\textbf{Gaussian fits to the HST/COS FUV emission line spectrum of the Leo~P \hii{} region.} The continuum-subtracted flux is plotted against the rest-frame wavelength (black line), and the best-fit Gaussian emission line model (Section~\ref{sec:fuv_fluxes}) is shown in purple.
While the continuum SNR ($\sim$3) in this region of the FUV spectrum is lower than in the optical, the O\,\textsc{iii}]\,$\lambda\lambda$1660, 1666 (left panel) and N\,\textsc{iii}]\,$\lambda$1749 (right panel) nebular emission lines are clearly detected. 
\label{fig:fuv_spectrum}}
\end{centering}
\end{figure*}

Figure~\ref{fig:fuv_spectrum} shows the region of the G160M HST/COS spectrum that covers the three nebular emission lines detected in the FUV: O\,\textsc{iii}]\,$\lambda\lambda$1660,\,1666 in the left panel, and N\,\textsc{iii}]\,$\lambda$1749 (actually two lines at 1748.65 and 1749.67\,\AA{}) in the right panel.  
As for the optical spectrum, we subtract a fifth-order polynomial fit to the continuum level to produce a pure nebular emission line spectrum.
The black line shows the continuum-subtracted flux as a function of wavelength in the rest frame of Leo~P.
The prominent absorption feature is a Milky Way ISM Al\,\textsc{ii}\,$\lambda$1670 line, which is blueshifted relative to the rest frame of Leo~P. 
The SNR of $\sim$3 in the continuum is lower in this part of the G160M spectrum than achieved in the KCWI data ($\sim$15), but the FUV nebular emission lines are still clearly detected.
It is surprising that we observe the N\,\textsc{iii}]\,$\lambda$1749 line in Leo~P, as it is rarely detected in metal-poor, star-forming galaxies locally or at high redshift \citep[e.g.,][]{stark14, berg22}.

We use a similar Gaussian fitting procedure to that described in Section~\ref{sec:optical_fluxes} to measure the emission line fluxes from the HST/COS spectrum.
We fit Gaussians to the O\,\textsc{iii}]$\,\lambda\lambda$1660.81,\,1666.15 and N\,\textsc{iii}]$\,\lambda\lambda$1748.65,\,1749.67 emission lines\footnote{There are five N\,\textsc{iii}] transitions between $1746-1754$\,\AA{}, but these two are expected to have the highest emissivity in \hii{} regions, and the N\,\textsc{iii}]$\,\lambda\lambda$1748.65,\,1749.67 pair is the only possibility that matches the wavelength spacing of the observed emission lines.}, requiring the same velocity FWHM for all lines because little change in the instrumental contribution is expected over this $\sim100$\,\AA{} wavelength range.
However, we find that these lines could not be modeled with a single velocity: while the best fit to the O\,\textsc{iii}] lines, $257.7\,\mathrm{km\,s}^{-1}$, agrees nicely with the systemic velocity of Leo~P (Table~\ref{tab:properties}), the N\,\textsc{iii}] lines are blueshifted by $-57.5\,\mathrm{km\,s}^{-1}$ relative to the O\,\textsc{iii}] lines.
The physical cause of the blueshifted N\,\textsc{iii}] emission is not clear, but we allow for this velocity offset in the N\,\textsc{iii}] lines in our modeling to enable good, simultaneous fits to all observed emission lines.
The best-fit emission line model is plotted as the purple line in Figure~\ref{fig:fuv_spectrum}, and we return to the anomalous N\,\textsc{iii}] emission in Section~\ref{sec:fuv_puzzles} below.

Again, we resample the observed, flux-calibrated spectrum 10,000 times and repeat our continuum normalization and Gaussian fitting procedure.
We find that the observational uncertainties on the COS spectrum were overestimated relative to the standard deviation in the featureless continuum regions, which is itself an upper limit on the true observational uncertainty due to the presence of weak metal absorption lines throughout the FUV continuum.
Consistent with our approach for the uncertainties on the KCWI data, we calculate the scale factor required to bring the median uncertainty into agreement with the observed continuum RMS.
We find that the COS flux uncertainties must be scaled down by a factor of three to ensure realistic uncertainties on our line flux measurements.
The uncertainty adopted for each line is half of the difference between the 84 and 16 percentile values of fluxes measured from the resampled spectra.
Table~\ref{tab:obs_lines} reports the measured fluxes for each component of the O\,\textsc{iii}] doublet, but only a total flux in both N\,\textsc{iii}] lines because the weaker line would not be formally detected at the 2-$\sigma$ level alone.
We correct those fluxes for extinction to calculate $I_\lambda$ by adopting the \ebv{} measured from the KCWI observations. 
Finally, we report the ratio of the weaker FUV line intensities to that of O\,\textsc{iii}]\,$\lambda$1666.

\subsection{Physical Conditions from Emission Line Ratios\label{sec:neb_conditions}}

From the ratios of emission line strengths measured from our KCWI observations, we can determine key properties of the \hii{} region.
The intrinsic (i.e., dust-free) ratios of the Balmer lines are well-known at a given nebular temperature in the limit of Case B recombination, where the gas is optically thick to photons more energetic than 13.6\,eV.
Thus, \ebv{} can be determined for an adopted dust extinction law from the observed ratio of H$\gamma$ or H$\delta$ to H$\beta$.

Similarly, the intrinsic ratio of the auroral \oiiiwave{4363} line to \oiiiwave{4959} or \oiiiwave{5007} is sensitive to the nebular temperature at a given electron density, $n_e$.  
Unfortunately, our KCWI spectra do not cover the most common nebular density diagnostic lines in the optical ([O\,\textsc{ii}]\,$\lambda\lambda$3727,\,3729 or [S\,\textsc{ii}]\,$\lambda\lambda$6717,\,6731).
We do detect the higher-ionization, density-sensitive [Ar\,\textsc{iv}]\,$\lambda\lambda$4711,\,4740 lines, but because this diagnostic saturates at low $n_e$, our [Ar\,\textsc{iv}]\,$\lambda$4711/[Ar\,\textsc{iv}]\,$\lambda4740$ measurement of $1.38 \pm 0.64$ only yields a rough upper limit of $n_e \lesssim 10^{2-3}$\,cm$^{-3}$ \citep[e.g.,][]{mingozzi22}.
Therefore, we adopt the electron density $n_e= 45^{+66}_{-45}$\,cm$^{-3}$ \citep{skillman13} determined from the [O\,\textsc{ii}] doublet in LBT/MODS long-slit spectroscopy of the Leo~P \hii{} region. 
The density is securely in the low-density limit, meaning that the temperature we derive from our observed \oiii{} lines is not sensitive to the precise $n_e$ that we adopt. 

Because we must correct the observed \oiii{} lines for reddening to determine the temperature, which sets the intrinsic Balmer line ratios, we adopt an iterative procedure.
We begin by assuming the nebular temperature $\mathrm{T(O}^{++})=17350$\,K \citep{skillman13} to calculate the intrinsic Case B Balmer line ratios using \pyneb{} \citep{luridiana12, luridiana15}.
We use the default atomic data in \pyneb{} for all calculations except those involving \oiii{}, for which we use the \citet{aggarwal99} collision strengths calculated from a six-level atom approximation required to analyze the FUV \oiiidoub{1660,\,1666} lines.
We then determine \ebv{} from the observed H$\gamma$/H$\beta$ and H$\delta$/H$\beta$ ratio compared to the intrinsic ratios, assuming a \citet{cardelli89} extinction law and $R_V=3.1$, and average the measurements from the two Balmer ratios. 
We find $E(B-V) = 0.0092$, and check that the results are not affected by using the average instead of only the ratio involving the stronger H$\gamma$ line (which gives $E(B-V) = 0.0086$).
This \ebv{} is then used to correct the observed \oiii{} lines for dust.
We then use the \pyneb{} routine \texttt{getTemDen} to infer \otemp{} from the dust-corrected \oiiiwave{4363}/\oiiiwave{4959} and \oiiiwave{4363}/\oiiiwave{5007} ratios, and average the two values.
This process is repeated using the updated \otemp{} as the starting point to determine the intrinsic Balmer line ratios until the \ebv{} and \otemp{} values change by less than 1\%.

To determine the uncertainties in these derived quantities, we redraw the emission line fluxes 10,000 times from Gaussians centered on the measured values and with $\sigma$ equal to the uncertainties on those line fluxes reported in Table~\ref{tab:obs_lines}.
We repeat the iterative process described above to measure \ebv{} and \otemp{} for all 10,000 samples and adopt half the 16-84 percentile range as the uncertainty on each measurement.

\begin{table}
\tabcolsep=0.5cm
\begin{center}
\caption{Physical Properties Derived from Emission Lines\label{tab:results}} 
\begin{tabular}{ll}
\hline
\multicolumn{2}{c}{\textbf{Leo~P KCWI Observations}} \\
$E(B-V)$ (mag) & $0.009 \pm 0.014$ \\
$\mathrm{T(O}^{++})$ (K) & $17320 \pm 120$ \\
\qh{} (s$^{-1}$) & $(3.74 \pm 0.67) \times 10^{48}$ \\
\qhei{} (s$^{-1}$) & $(4.42 \pm 0.80) \times 10^{47}$  \\
\qhei{}/\qh{} & $0.12 \pm 0.03$ \\
$\xi_\mathrm{ion}$ (s$^{-1}$\,/\,erg\,s$^{-1}$\,Hz$^{-1}$) & $(2.68 \pm 0.55) \times 10^{25}$ \\
\multicolumn{2}{c}{\textbf{Model Fit to SED of LP26}} \\
\qh{} (s$^{-1}$) & $4.25 \times 10^{48}$ \\
\qhei{} (s$^{-1}$) & $7.80 \times 10^{47}$  \\
\qhei{}/\qh{} & 0.18 \\
\end{tabular}
\end{center}
\tablecomments{\ebv{} and \otemp{} are the nebular reddening and temperature measured in Section~\ref{sec:neb_conditions}. \qh{} and \qhei{} are the production rate of H-ionizing and He-ionizing photons, calculated from the observed \hii{} region emission in Leo~P. $\xi_\mathrm{ion}$ is the H-ionizing photon production efficiency, i.e., the ratio of \qh{} to the luminosity density at 1500\,\AA{}. For comparison, we also report \qh{} and \qhei{} predicted by the best-fit model to the observed FUV-through-NIR SED of LP26, the ionizing O star, from \citet{telford21}.}
\end{table}

Table~\ref{tab:results} reports the properties of the Leo~P \hii{} region measured from the KCWI spectra.
The \ebv{} that we infer is comparable to the foreground $E(B-V) = 0.014$ mag toward Leo~P \citep{green15}, implying low internal extinction consistent with the very low metallicity of Leo~P (3\%\,\zsun{}; \citealt{skillman13}).
As described above, we use this \ebv{} measurement and the \citet{cardelli89} extinction law to correct the observed emission line fluxes for dust extinction, and report the resultant ratios $I_\lambda/I_{\mathrm{H}\beta}$ (or $I_\lambda/I_{\lambda 1666}$ for the FUV lines) in Table~\ref{tab:obs_lines}.

The \otemp{} derived here is very similar to that reported by \citet{skillman13}, though we note that the values we calculate from the two \oiii{} ratios differ: the \oiiiwave{4363}/\oiiiwave{4959} and \oiiiwave{4363}/\oiiiwave{5007} ratios imply $17150 \pm 100$\,K and $17490 \pm 110$\,K, respectively. 
We have checked that adopting any of these \otemp{} values negligibly changes either the derived \ebv{} (3\% change) or calculations presented in Section~\ref{sec:results} below ($<$\,1\% change).
Perhaps relatedly, we find that the observed ratio of the strong \oiii{} line fluxes, \oiiiwave{5007}/\oiiiwave{4959} = $3.00 \pm 0.01$ (consistent with the ratio reported in \citealt{skillman13}), differs from the theoretical ratio of 2.89 at these nebular temperatures predicted by \pyneb{}. 
The theoretical ratio is highly sensitive to the adopted atomic data, and using a different set of collision strengths for \oiii{} could easily bring the theoretical value closer to the ratio we observe. 
Again, our choice of atomic data is limited by the need for six-level atom calculations to analyze the FUV O\,\textsc{iii}] lines.


\section{Results: Empirical Constraints on the Ionizing Spectrum of an Extremely Metal-Poor O Star\label{sec:results}}

\subsection{Total Ionizing Flux from LP26\label{sec:qion}}

With our measurements of integrated line intensities for the entire \hii{} region and physical conditions of the nebula in hand, we are in a position to calculate the ionizing flux from the central O star, LP26, that is necessary to produce the observed nebular emission in Leo~P.
The H-ionizing photon production rate, \qh{}, is related to the H$\beta$ intensity by:
\begin{equation} \label{eq:qion}
Q(\mathrm{H}) = \frac{4 \pi D^2}{\left(1 - f_\mathrm{esc}\right)} \, \frac{I_{\mathrm{H}\beta}}{h \nu_{\mathrm{H}\beta}} \, \frac{\alpha_\mathrm{B}}{\alpha^\mathrm{eff}_{\mathrm{H}\beta}},
\end{equation}
where $D$ is the distance to Leo~P (Table~\ref{tab:properties}), \fesc{} is the fraction of ionizing photons that escape the \hii{} region, $h \nu_{\mathrm{H}\beta}$ is the energy of a H$\beta$ photon, and $\alpha_\mathrm{B}$ and $\alpha^\mathrm{eff}_{\mathrm{H}\beta}$ are the total H and effective H$\beta$ Case B recombination coefficients.
We calculate the recombination coefficients at 17510\,K (Table~\ref{tab:results}; determined from the observed [O\,\textsc{iii}] line ratios in Section~\ref{sec:neb_conditions} above) using the fitting formulae from \citet{pequignot91}.
Because the \hii{} region is isolated and spherical, similar to an ideal Str{\"o}mgren sphere, we adopt \fesc{} = 0, assuming that all ionizing photons interact within the \hii{} region.
From the observed emission line fluxes, we calculate $Q(\mathrm{H}) = (3.74 \pm 0.67) \times10^{48}\,\mathrm{s}^{-1}$.
All ionizing photon production calculations are reported alongside other physical properties derived from the observed optical nebular emission lines in Table~\ref{tab:results}.

Again, the formal uncertainty on \qh{} is calculated by repeating these calculations for the 10,000 resampled emission line fluxes (and resampled distances to Leo~P) and adopting half the 16-84 percentile range; this is the value reported in Table~\ref{tab:results}.
However, there is also systematic uncertainty in the choice of $f_\mathrm{esc}=0$. 
Multiple lines of reasoning support this assumption: first, there is no evidence for a porous \hii{} region structure that would suggest significant leakage of ionizing photons into the surrounding ISM; and 
second, we detect clear [S\,\textsc{ii}] lines that arise in the low-ionization, outer part of the nebula, implying that the \hii{} region is ionization-bounded (i.e., optically thick to ionizing photons; \citealt{pellegrini11, wang21}).
While the available data point to a low \fesc{} close to zero, ultimately we are measuring a lower limit on \qh{} and the true value may be higher if a significant fraction of ionizing photons escape the \hii{} region.

Dust may absorb ionizing photons, making them unavailable to interact with the gas and causing the \qh{} implied by the observed H$\beta$ emission to be lower than the true value.
However, our \ebv{} measurement here and previous observational estimates of internal extinction in the \hii{} region all suggest $A_V < 0.1$ \citep{skillman13, telford21}.
Thus, absorption of ionizing photons by dust should not be significant.

Next, we turn to the question of how our \qh{} measurement for LP26 compares to the predictions of theoretical stellar spectral models at extremely low metallicity.
We compare to the \qh{} predicted by the best-fit model to the spectral energy distribution (SED) of LP26 reported in \citet{telford21}. 
In that work, \tlusty{} \citep{hubeny95} spectral models from the \textsc{ostar2002}\footnote{\url{http://tlusty.oca.eu/Tlusty2002/tlusty-frames-OS02.html}} grid \citep{lanz03} were combined with \textsc{parsec}\footnote{\url{http://stev.oapd.inaf.it/cgi-bin/cmd_3.1}} stellar evolution models \citep{bressan12}.
The latter are required to match appropriate stellar radii, which set the bolometric luminosity and normalization of the stellar SED, with the spectral models at each combination of \teff{}, \logg{} in the \textsc{ostar2002} model grid.
The best-fit model not only reproduces the overall FUV-through-NIR SED shape, but nicely matches the observed photospheric line profiles in the FUV spectrum, lending confidence to its ability to match metal line opacities relevant for the ionizing photon production.

By simply integrating the best-fit SED model over all energies above the 13.6\,eV required to ionize H, we find a prediction of $Q(\mathrm{H}) = 4.25\times10^{48}\,\mathrm{s}^{-1}$.
This is 13.6\% higher than, but consistent within the uncertainties with, the \qh{} inferred from the H$\beta$ emission (and measured \ebv{} and \otemp) in the KCWI observations of the \hii{} region powered by LP26.
This is remarkably good agreement given the lack of empirical constraints on stellar spectral and evolution models at such low metallicities.

This \qh{} value normalized to the UV continuum luminosity density ($L_{1500}$, in erg\,s$^{-1}$\,Hz$^{-1}$) measured from the HST/COS spectrum of LP26 (Section~\ref{sec:cos_data}) gives the star's ionizing photon production efficiency, $\xi_\mathrm{ion}$. 
We calculate $\log \xi_\mathrm{ion} = 25.4$ for LP26, similar to estimates for massive stellar populations at high $z$ (e.g., \citealt{hutchison19, endsley21}), and higher than the value of 25.2 required for star-forming galaxies to reionize the early universe with a moderate escape fraction (5-20\%; e.g., \citealt{ouchi09, robertson15, finkelstein19, naidu20}).

The model-predicted \qh{} depends on several stellar properties, particularly its bolometric luminosity (\lstar).
The \textsc{ostar2002} spectra are given in terms of the Eddington flux at the stellar surface, which must be scaled by $4\pi R_\star^2$ to obtain the total number of photons emitted by the star per second, per frequency. 
Thus, each model in the \textsc{ostar2002} grid is not associated with a unique \qh{}.
Even if the best-fit \teff{} of LP26 from the SED fitting is accurate, detailed stellar atmosphere modeling of the star's FUV and optical spectra (Telford et al.\ 2023, in preparation) may improve upon our current estimates of the surface gravity and luminosity.
Future work will assess whether a spectral model produced with a modern code and tailored to the observed photosphere and wind features brings the predicted \qh{} into even better agreement with the KCWI observations.

The possible binary status of LP26 would also impact the \qh{} measured from nebular emission. 
Most massive stars are thought to begin their lives with at least one companion \citep[e.g.,][]{sana12, offner22}.
No spectroscopic evidence for a binary companion (e.g., radial velocity variations, complex absorption line profiles, or photospheric features suggestive of significantly different \teff{}) currently exists for this star, but that does not rule out the possibility of a companion.
A roughly equal-mass binary companion would double the system's \lstar{} and \qh{} over expectations for a single star. 
But this is inconsistent with the good agreement we find between the single-star SED model fit to observations of LP26 and the \qh{} measured from the \hii{} region emission that it powers.
In principle, an equal-mass binary combined with $f_\mathrm{esc}\approxeq50\%$ could conspire to mimic a single star, but such a high \fesc{} is unlikely given the strong, low-ionization [S\,\textsc{ii}] emission in the KCWI data that we discussed above.
At present, there is no compelling evidence that LP26 has a binary companion, so the simple, single-star interpretation is appropriate.
Given the many factors that impact \qh{}, our finding that nebular emission is consistent with the SED model at the $\sim$\,10\% level is a great success for the current generation of stellar models at extremely low $Z$.

Whether \qh{} depends on $Z$ at fixed SpT remains unclear, as known variations in other stellar properties with $Z$ have opposite impacts on \qh{}.
More metal-poor stars are known to have higher \teff{} at fixed SpT due to weaker ``backwarming'' from metal line blanketing \citep{martins05}, but a zero-age main-sequence star with a given \teff{} will be less massive and less luminous at lower $Z$ \citep{evans19}.
There is also considerable scatter in the relationship between \qh{} and SpT: our measured \qh{} is consistent with the range for O7$-$B0 SpTs from models fit to observations of stars in the Magellanic Clouds \citep{ramachandran19}.
At higher metallicity, \citet{martins05} predicted \qh{} as a function of spectral type (SpT) and luminosity class for Galactic O stars.
Comparing to their results for luminosity class V, we find that the $\log{Q(\mathrm{H})} = 48.6$ measured here (both from KCWI data and from the SED model) corresponds to an O7.5 SpT -- consistent with the O7$-$8~V estimated for LP26 based on HST photometry (\citealt{mcquinn15f}; strong \hei{} emission in the KCWI data prevents a robust spectroscopic SpT determination).
Additional empirical constraints on \qh{} for a larger sample of very metal-poor O stars are required to determine how \qh{} varies with $Z$.


\subsection{The Shape of the Ionizing Spectrum\label{sec:spectral_shape}} 

In the previous section, we calculated the production rate of all photons with energies greater than 13.6\,eV from the \hbeta{} intensity, which gives the normalization of LP26's ionizing spectrum.
Yet, we have measured the intensities of many nebular emission lines of ions with a wide range of ionization energies.
Here, we use this rich dataset to constrain the relative production of photons above different energy thresholds; that is, the shape of the ionizing spectrum of LP26.

\subsubsection{Helium-Ionizing Flux\label{sec:q_ratios}}

Analogously to the calculation of \qh{} in Section~\ref{sec:qion}, the production rate of photons capable of ionizing He (with energies above 24.6\,eV) can be calculated from the observed intensity of the He\,\textsc{i}\,$\lambda$4471 line. 
We determine \qhei{} using equation~\ref{eq:qion} but with the intensity, frequency, and total and effective recombination rates (from the equations of \citealt{pequignot91}) appropriate for the He\,\textsc{i}\,$\lambda$4471 line, instead of for \hbeta{}. 
Again, we assume that no photons escape the nearly spherical, apparently undisturbed \hii{} region.
This calculation gives $Q(\mathrm{He}) = (4.42 \pm 0.80)\times10^{47}\,\mathrm{s}^{-1}$ produced by LP26. 

The ratio of \qhei{}/\qh{} is a highly useful diagnostic of the hardness of the ionizing spectrum, particularly because this quantity is not affected by uncertainties in the absolute flux calibration of the observations or in comparing the bolometric luminosity of the ionizing source to stellar models.
Dust should also have a negligible impact, given the very low \ebv{} we measure from the Balmer lines in this extremely metal-poor galaxy. 
From the KCWI observations of the Leo~P \hii{} region, we find $Q(\mathrm{He})/Q(\mathrm{H}) = 0.12 \pm 0.03$ for LP26.

\begin{figure}[!tp]
\begin{centering}
  \includegraphics[width=\linewidth]{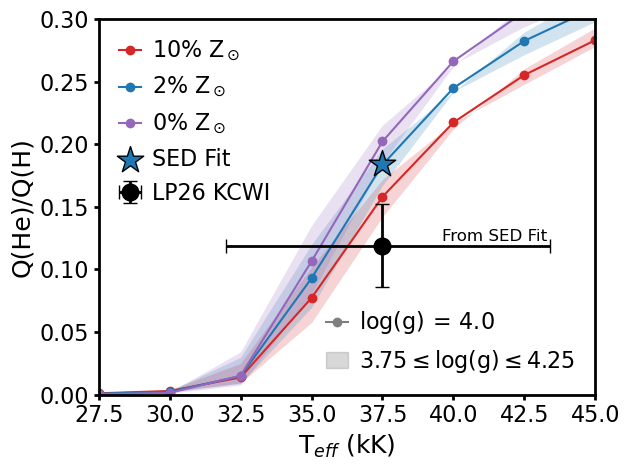}
\caption{\textbf{Comparing $\bm{Q(\mathrm{He})/Q(\mathrm{H})}$ of LP26 measured from nebular emission to \tlusty{} models.} \qhei{}/\qh{} is plotted as a function of \teff{}, with red, blue, and purple lines showing models with $\log(g) = 4.0$ and $Z=10, 2,$ and 0\%\,\zsun{}, respectively. The blue star indicates the best-fit \tlusty{} model to the SED of LP26 reported in \citet{telford21}, and the shaded regions show the impact of varying \logg{} between 3.75 and 4.25, roughly the range allowed by that modeling.
The black circle with error bars indicates the \qhei{}/\qh{} measured from the KCWI observations in Leo~P and the \teff{} of LP26 from the SED fit.
The observed H$\beta$ and He\,\textsc{i}\,$\lambda$4471 emission suggests that the O star's true \teff{} lies between 35 and 37.5\,kK.
\label{fig:qion_vs_teff}}
\end{centering}
\end{figure}

Again, we compare to the prediction of the best-fit \tlusty{} spectral model for LP26 reported in \citet{telford21}: $Q(\mathrm{He})/Q(\mathrm{H}) = 0.18$. 
The slight mismatch implies that the 2\%\,\zsun{} (relative to $Z=0.0017$; \citealt{grevesse98}), $T_\mathrm{eff}=37.5$\,kK model spectrum is harder than that required to produce the nebular emission we observe with KCWI.
The \textsc{ostar2002} grid is somewhat coarsely sampled in \teff{} steps of 2.5\,kK; the predicted \qhei{}/\qh{} is 0.09 for the next-lowest $T_\mathrm{eff}=35$\,kK in the 2\%\,\zsun{} model grid.
This value is lower than the observed ratio, suggesting that the \teff{} of LP26 is intermediate between those two values (but still within the reported uncertainty on the best-fit \teff{} reported in \citealt{telford21}). 
Again, the \qhei{}/\qh{} determined from the KCWI observations is very close to the value of 0.107 reported for SpT O7.5~V in the Milky Way (\citealt{martins05}; for comparison, that paper reports $Q(\mathrm{He})/Q(\mathrm{H}) = 0.135$ for O7~V and 0.072 for O8~V).

Figure~\ref{fig:qion_vs_teff} presents a comparison of the observational constraints on the \teff{} and hardness of the ionizing spectrum of LP26 with the predictions of \tlusty{} models for a range of \teff{}, \logg{}, and metallicity. 
\qhei{}/\qh{} is plotted as a function of \teff{}, with red, blue, and purple lines showing models with $Z=10, 2,$ and 0\%\,\zsun{}, respectively.
For reference, the blue star indicates the best-fit \tlusty{} model reported in \citet{telford21}, and the range of \logg{} shown corresponds to the uncertainties from the SED fitting.
The black circle with error bars indicates the best-fit \teff{} from modeling the SED of LP26 and the \qhei{}/\qh{} measured from the KCWI observations in Leo~P.

From Figure~\ref{fig:qion_vs_teff}, it is clear that the \qhei{}/\qh{} ratio is primarily a function of \teff{}, with a weaker dependence on both \logg{} and metallicity. 
The \qhei{}/\qh{} of the metal-poor \tlusty{} models are fully consistent with the observations within the uncertainties.
Clearly, the \teff{} inferred from SED fitting is rather uncertain; tighter constraints will be possible from fitting new atmosphere models to the KCWI observations of LP26's stellar continuum features (Telford et al.\ 2023, in preparation). 
We emphasize that this precision is limited in part by the 2.5\,kK spacing between \teff{} values in the \textsc{ostar2002} grid, and that a more finely sampled grid of spectral models would likely produce a better fit and more precise \teff{} inference.
The narrow range of \qhei{}/\qh{} allowed by the KCWI observations implies a slightly lower \teff{} for LP26 than inferred from the best-fit SED model, but within one step in the \tlusty{} grid (i.e., between 35 and 37.5\,kK) and within the reported uncertainty on \teff{}. 
We have found a remarkable level of consistency across various observational constraints on the properties of LP26, suggesting that the \tlusty{} model spectra perform well in this very low-$Z$ regime.

We do not detect He\,\textsc{ii}\,$\lambda$4686 nebular emission in the KCWI spectrum, nor He\,\textsc{ii}\,$\lambda$1640 in the HST/COS data. 
The higher-quality KCWI observations give a more stringent 2-$\sigma$ upper limit on the He\,\textsc{ii}\,$\lambda$4686 flux of $1.76\times10^{-18}$\,erg\,cm$^{-2}$\,s$^{-1}$.
This implies a very low production rate of photons capable of ionizing He$^+$ (with energies above 54.4\,eV) of $Q(\mathrm{He}^+) \leq 7.29\times10^{44}\,\mathrm{s}^{-1}$ (or $Q(\mathrm{He}^+)/Q(\mathrm{H}) \leq 1.95\times10^{-4})$, consistent with the expectation only the hottest O stars ($T_\mathrm{eff} \gtrsim 80$\,kK) produce enough of these highly energetic photons to drive strong He\,\textsc{ii} nebular emission \citep[e.g.,][]{schaerer02}.
For reference, the model fit to LP26's SED predicts $Q(\mathrm{He}^+) = 2.3\times10^{41}\,\mathrm{s}^{-1}$ and $Q(\mathrm{He}^+)/Q(\mathrm{H}) = 5.62\times10^{-8}$.

\subsubsection{\textsc{Cloudy} Photoionization Modeling\label{sec:cloudy}}

\begin{figure*}[!htp]
\begin{centering}
  \includegraphics[width=\linewidth]{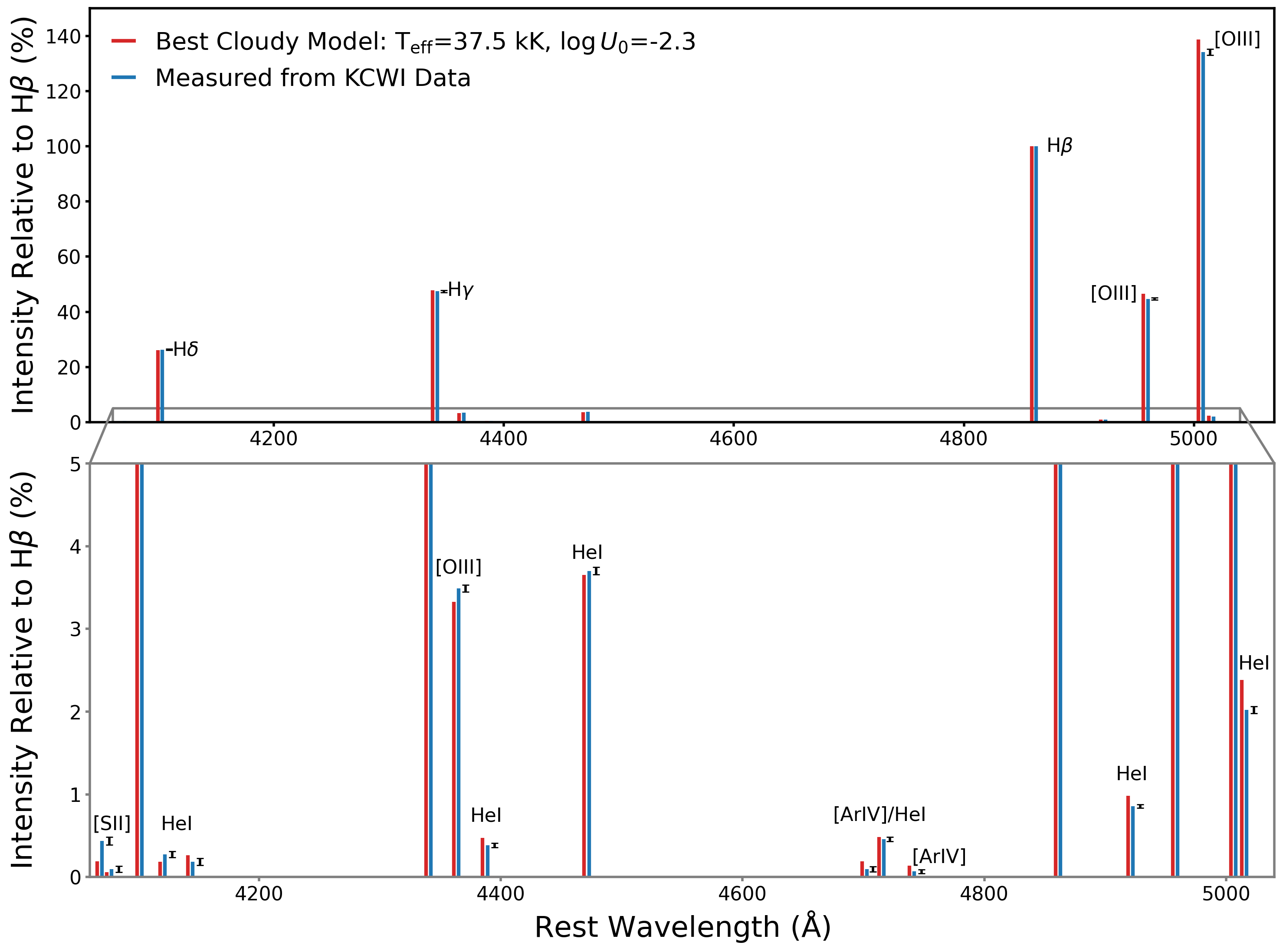}
\caption{\textbf{The spectral model fit to the observed SED of LP26 reproduces the observed nebular emission.} \underline{Top}: Vertical red and blue lines represent the emission line intensities relative to H$\beta$ in the best-fit \cloudy{} model and measured from the KCWI data, respectively. 
Black error bars show the uncertainties on the line intensities measured from the KCWI observations (blue lines); these do not account for uncertainty in the small extinction correction, but the excellent agreement between observed and model Balmer line ratios with respect to H$\beta$ suggest that correction is robust.
Each line's intensity is plotted at its rest wavelength with a small offset for clarity. 
The grey rectangle indicates the region shown in more detail in the bottom panel.
\underline{Bottom}: Zoom-in to the region within the grey rectangle in the top panel to highlight the weak lines ($<$\,5\% of the H$\beta$ intensity). 
The \cloudy{} model with a 37.5\,kK ionizing spectrum is able to match simultaneously nearly all of the observed emission line strengths, which require photons with a wide range of ionization energies to produce.
\label{fig:cloudy_optical}}
\end{centering}
\end{figure*}

\begin{figure}[!tp]
\begin{centering}
  \includegraphics[width=\linewidth]{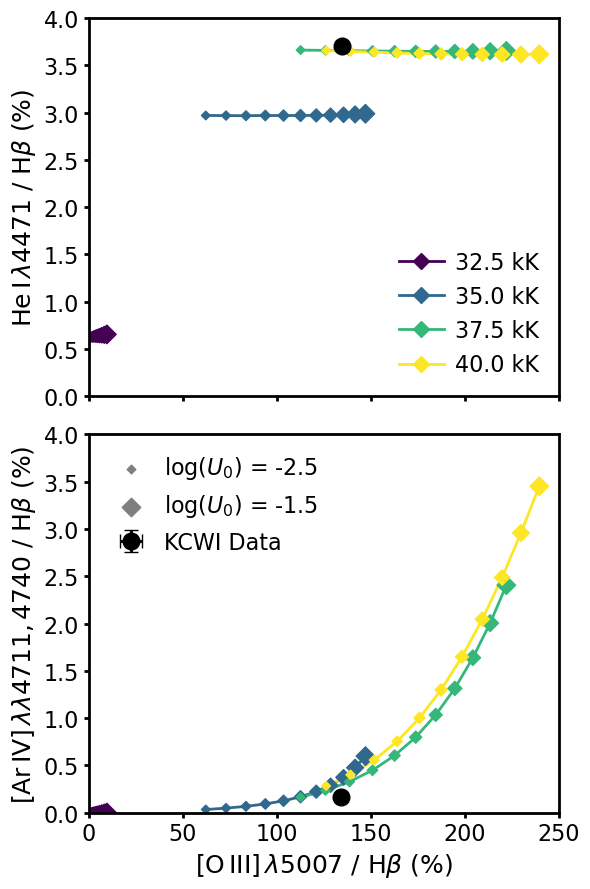}
\caption{\textbf{Dependence of \cloudy{}-predicted emission line strengths on $\bm{T_\mathrm{eff}}$ and $\bm{\log{U_0}}$.} He\,\textsc{i}\,$\lambda$4471/H$\beta$ (top) and [Ar\,\textsc{iv}]\,$\lambda\lambda$4711,\,4740/H$\beta$ (bottom) are plotted as a function of [O\,\textsc{iii}]\,$\lambda$5007/H$\beta$.
The diamonds connected by solid lines show predictions of the \cloudy{} models for a given \teff{}, and increasing point size indicates increasing $\log{U_0}$ (from left to right). 
Colors encode the varying \teff{} of the \tlusty{} stellar spectrum adopted as the ionizing source in each model: purple, blue, green, and yellow correspond to 32.5, 35.0, 37.5, and 40.0\,kK, respectively. 
The emission line flux ratios measured from the KCWI observations are shown as black circles with error bars (which are smaller than the point size). 
This 37.5\,kK model spectrum is nearly able to match simultaneously all of these high-ionization lines. 
\label{fig:cloudy_lineratios}}
\end{centering}
\end{figure}

In addition to the relatively simple H and He lines, we have detections of several collisionally excited metal lines. 
Since these are not recombination lines, we cannot simply calculate the production rate of photons capable of ionizing the metal species, but the relative strengths of these various lines place strong constraints on the shape of the ionizing spectrum.
As a complementary test of the ionizing star's \teff{} to the analysis in Section~\ref{sec:q_ratios} above, we compare our observed emission line ratios to the predictions of photoionization modeling.

We use the photoionization code \cloudy{} \citep[v.17.03;][]{ferland98, ferland17} to model the nebular emission from the \hii{} region in Leo~P.
We adopt the gas-phase metal abundances reported in \citet{skillman13}, and use the updated measurement of He abundance from \citet{aver22}. 
For all other metal species for which abundances have not been measured in Leo~P, we use the GASS10 \citep{grevesse10} solar abundance set provided in \cloudy{}, scaled to 3\%\,\zsun{} (chosen to match the abundance of oxygen relative to solar), and account for depletion of metals onto dust grains.
The \hii{} region is assumed to have a plane-parallel geometry and both covering and filling factors of unity.
Because H is fully ionized, we assume a constant total H density, $n_\mathrm{H}$, equal to $n_e = 45\,\mathrm{cm}^{-3}$ \citep{skillman13}. 

The predicted nebular line emission is sensitive to the geometry of the \hii{} region, specifically, its size (i.e., the distance from the ionizing source to the inner face of the ionized cloud, $R_0$) and the gas distribution and density. 
Because our goal is to constrain the shape of the star's ionizing spectrum, we simplify the parameter space that we explore (following, e.g., \citealt{byler17}) by combining these uncertain and degenerate \hii{} region properties into the dimensionless ionization parameter $U_0$.
In a plane-parallel geometry and for a fixed ionizing spectrum, the line emissivities will be identical for any combination of \qh{}, $n_\mathrm{H}$, and $R_0$ that gives the same value of $U_0$:
\begin{equation}
U_0 = \frac{1}{4 \pi \, c}\,\frac{Q(\mathrm{H})}{R_0^2 \, n_\mathrm{H}}.
\end{equation}
Given model ionizing stellar spectra with varying \teff{} and the fixed \hii{} region properties described above, we search over $-2.5 \leq \log U_0 \leq -1.5$, typical of normal star-forming galaxies, to find the best match to the observed nebular emission in Leo~P.

We use \tlusty{} model spectra from the \textsc{ostar2002} grid, which are implemented in \cloudy{}.
As illustrated in Figure~\ref{fig:qion_vs_teff}, the relative strength of emission from various ions depends more strongly on the model \teff{} than on either \logg{} or $Z$, so we fix \logg{}\,$=$\,4.0 and $Z$=$1/50$\,\zsun{}, following the best-fit model to the SED of LP26 reported in \citet{telford21}. 
We run \cloudy{} simulations for $T_\mathrm{eff} = \{32.5, 35.0, 37.5, 40.0\}$\,kK, spanning the values of \teff{} preferred by SED modeling and by the \qhei{}/\qh{} measured from KCWI observations in Section~\ref{sec:q_ratios} above, and also sampling an additional step in the \teff{} grid above and below for completeness.
We vary $\log U_0$ in steps of 0.1 for each \teff{} value and predict the intensities (neglecting dust extinction) relative to H$\beta$ of all nebular emission lines detected in the KCWI data.
These line intensity ratios are directly comparable to the dust-corrected $I_\lambda$/$I_{\mathrm{H}\beta}$ reported in Table~\ref{tab:obs_lines}.

We perform a $\chi^2$ minimization to determine the best overall fit to the observed optical emission line spectrum; the FUV lines are not included in this optimization.
The best agreement between the \cloudy{} predictions and KCWI observations is found for $T_\mathrm{eff} = 37.5$\,kK and $\log U_0 = -2.3$, which gives a total $\chi^2 = 236.7$ (vs.  $\chi^2 = 257.7$ and 400.0 for the best model at 40 and 35\,kK, respectively).
Figure~\ref{fig:cloudy_optical} shows the comparison between the observed optical nebular emission line intensities and the best-fit \cloudy{} model.
The heights of the vertical lines represent the emission line intensities with respect to H$\beta$ as measured from the KCWI data (blue) and as predicted by \cloudy{} (red).
Each line is plotted at the rest wavelength of the transition, with a few-\AA{} offset for clarity.
Black error bars next to the blue lines show the measurement uncertainties.
The uncertainty in the correction for dust extinction is not included in those error bars, but the measured \ebv{} is small and the dust-corrected observed Balmer line ratios agree very nicely with the model predictions, to which no dust extinction has been applied. 
The strongest lines are visualized in the top panel, while the weaker lines enclosed in the grey rectangle in the top panel are shown in more detail in the bottom panel.

Overall, the agreement between observed and predicted nebular emission line intensities with respect to H$\beta$ is remarkably good. 
The observed strengths of the Balmer lines, temperature-sensitive [O\,\textsc{iii}] lines, and high-ionization [Ar\,\textsc{iv}] lines (which require photons with energies $\geq 40.7$\,eV) are all simultaneously reproduced by the best model within the 5-$\sigma$ level.
This is excellent agreement, particularly considering that the simple geometry adopted in any photoionization model cannot capture the complex distribution of gas in real star-forming regions.
Most He\,\textsc{i} lines also agree nicely, particularly the 4471\,Å line that we used to constrain the He-ionizing photon production in Section~\ref{sec:q_ratios} above.
Slight disagreement in some He\,\textsc{i} lines may be due to uncertainties in atomic data or nebular density structure, or even stellar absorption features that are difficult to constrain because they are nearly completely filled in by the nebular emission.
The latter effect should be small, but will have greater impacts on the weaker He\,\textsc{i} lines.

Figure~\ref{fig:cloudy_lineratios} presents a comparison between our \cloudy{} model grid and the high-ionization emission lines (normalized to H$\beta$) measured from the KCWI data. 
He\,\textsc{i}\,$\lambda$4471/H$\beta$ (top) and [Ar\,\textsc{iv}]\,$\lambda\lambda$4711,\,4740/H$\beta$ (bottom) are plotted as a function of [O\,\textsc{iii}]\,$\lambda$5007/H$\beta$.
Diamonds connected by lines show predictions for an ionizing stellar spectrum of a fixed \teff{}, indicated by the color, and point size encodes the model $\log{U_0}$.
The observed, extinction-corrected emission line ratios measured from the KCWI observations are shown as black circles in each panel, and error bars showing the 1-$\sigma$ observational uncertainties are smaller than the points.
The He\,\textsc{i}\,$\lambda$4471 emission is strongly a function of \teff{} and saturates around 37.5-40\,kK, while the [Ar\,\textsc{iv}] and [O\,\textsc{iii}] emission both depend strongly on $\log U_0$.
Overall, the 37.5\,kK \tlusty{} model predicts emission line strengths that nearly match simultaneously all of these high-ionization transitions at a single $\log U_0$, suggesting that the hardness of the model spectrum is similar to that of LP26.

We showed in Section~\ref{sec:q_ratios} above that the observed \otemp{}, \ebv{}, and H$\beta$ and He\,\textsc{i}\,$\lambda$4471 emission imply a \qhei{}/\qh{} that is intermediate between the 35\,kK and 37.5\,kK \tlusty{} models. 
The top panel of Figure~\ref{fig:cloudy_lineratios} shows that \cloudy{} simulations for the 35\,kK stellar model predict substantially weaker He\,\textsc{i}\,$\lambda$4471 emission than observed.
That ionizing spectrum also requires $\log U_0=-1.7$ to match the [O\,\textsc{iii}] emission, which results in a poorer match to both the [Ar\,\textsc{iv}] and [S\,\textsc{ii}] doublets. 
Thus, a 35 kK stellar ionizing spectrum produces too few photons capable of producing He$^+$ to match the observations, in agreement with our analysis of the H$\beta$ and He\,\textsc{i}\,$\lambda$4471 lines alone.
Together, these results suggest that a \teff{} intermediate between the 35 and 37.5\,kK steps in the \tlusty{} model grid can explain the observed nebular emission from the Leo~P \hii{} region. 

The only notable mismatch in the best-fit model is in the low-ionization [S\,\textsc{ii}]\,$\lambda$4068 line, for which the model prediction is 5-$\sigma$ lower than the observed flux.
We found that none of the \cloudy{} models we tested was able to simultaneously match the high- and low-ionization lines.
The density structure of the \hii{} region is unlikely to be the culprit, as the $n_e$ measured by \citet{skillman13} from the [S\,\textsc{ii}]\,$\lambda\lambda$6717,\,6731 doublet was consistent with the low value measured from the [O\,\textsc{ii}] lines.
B-type stars in the \hii{} region could potentially produce excess photons capable of ionizing S ($\geq$\,10.4\,eV) without contributing to \qh{}, but this is unlikely because such stars must be much fainter than the O star (over 2 mag fainter in the F475W filter; \citealt{mcquinn15f}), implying a negligible contribution to the stellar spectrum at any wavelength.
On the other hand, photoionization modeling of [S\,\textsc{ii}] is notoriously difficult, as low-temperature dielectronic recombination rates have not been calculated for sulfur \citep[e.g.,][]{ali91, hill14}.
Aside from this small discrepancy, the \cloudy{} model is able to reproduce the observed emission line ratios, implying that the 37.5\,kK \tlusty{} model is a reasonable approximation of LP26's ionizing spectrum.


\section{Discussion\label{sec:discussion}}

\subsection{Implications for Modeling the Ionizing Fluxes of Extremely Metal-Poor Stellar Populations\label{sec:implications}}

Our constraints on the shape and normalization of the ionizing spectrum of LP26 from observed optical nebular emission (Section~\ref{sec:results} above) agree quite nicely with previous estimates of the star's \teff{} and \lstar{} based on its FUV spectrum and optical/NIR HST photometry \citep{telford21}.
The observed H$\beta$ and He\,\textsc{i}\,$\lambda$4471 emission, combined with measurements of \otemp{} and \ebv{} from the Balmer and [O\,\textsc{iii}] lines, are consistent with a theoretical \tlusty{} stellar spectrum with \teff{} and bolometric \lstar{} similar to, but slightly lower than, the values inferred  from the best fit to the SED of LP26. 
The \qhei{}/\qh{} ionizing spectral hardness diagnostic and the full \cloudy{} photoionization modeling both show that the observed nebular emission can be explained by a spectral model within one \teff{} grid step (i.e., between 35 and 37.5\,kK).
This is the first empirical validation of theoretical stellar spectra at significantly lower $Z$ than the 20$-$50\%\,\zsun{} of the Magellanic Clouds.

In this extremely low $Z$ regime, very few grids of theoretical spectra calculated with sophisticated stellar atmosphere codes are available. 
The \tlusty{} \textsc{ostar2002} grid used in this work (and to model the SED of LP26 in \citealt{telford21}) is  widely adopted for analyses of metal-poor individual stars and stellar populations, thanks to its comprehensive $Z$ sampling (including $1/10$, $1/30$, $1/50$, $1/100$, and $1/1000$\,\zsun{}, as well as metal-free models).
Of all the stellar atmosphere model grids that have been implemented in \cloudy{} (including WM-Basic and CoStar models; \citealt{pauldrach01} and \citealt{schaerer97}, respectively), only the \tlusty{} grid extends below 20\%\,\zsun{}.
However, these theoretical models have one important drawback: they do not include the effects of an expanding stellar wind. 

As ionizing photons absorbed by metal transitions in the stellar atmosphere are a key driver of stellar mass loss, this is highly relevant to accurately predicting the ionizing spectra of hot stars.
Dense stellar winds interact with ionizing photons through complex mechanisms \citep[e.g.,][]{schaerer97, martins21}, intricately connecting the stellar outflow with the emergent ionizing spectral shape.
Because metal transitions launch these radiation-driven winds, there is broad theoretical agreement that \mdot{} decreases toward lower $Z$, and winds may become negligible for extremely metal-poor stars \citep[e.g.,][]{vink01, bjorklund21, martins21}
The excellent consistency we find between the \tlusty{} model for LP26 and the nebular emission from its \hii{} region suggests that wind-free stellar ionizing spectra are appropriate for typical (i.e., moderate \mstar{} and \teff{}) O stars at 2-3\%\,\zsun{}.
Of course, additional observations of other extremely metal-poor, single-star \hii{} regions are needed to confirm that interpretation.

At higher $Z$, stronger stellar winds complicate theoretical ionizing spectra and must be treated properly in stellar atmosphere models to produce accurate density structures and emergent ionizing spectra \citep[e.g.,][]{sellmaier96}.
The various atmosphere codes in use take different computational approaches to modeling the hydrodynamics and radiative transfer in expanding stellar atmospheres, which impacts the ionizing spectra they predict \citep[e.g.,][]{simon-diaz08}.
At 50\%\,\zsun{}, \citet{zastrow13} found that spectra produced by codes that include stellar mass loss were better able to match the observed emission across $\sim$\,10 single-star \hii{} regions in the LMC than the softer ionizing spectra predicted by the \tlusty{} \textsc{ostar2002} grid, demonstrating that stellar winds do impact the ionizing spectra of O stars even below Galactic $Z$.

At present, very few published spectral model grids that include stellar wind physics extend below 20\%\,\zsun{}, and those that do \citep[e.g.,][]{martins21} are not commonly used either in \cloudy{} or in SPS codes.
On the observational side, we are limited to just one single-star \hii{} region at 3\%\,\zsun{}, which is not sufficient to determine which available models best represent ionizing spectra across the range of OB spectral types at extremely low $Z$.
Thus, we cannot at this stage determine whether it is important to include stellar mass loss to accurately predict the ionizing spectra of metal-poor galaxies; that will be the subject of a future paper (Telford et al.\ 2023, in preparation). 
The favorable comparison between the \qh{} and hardness of the ionizing spectrum inferred from our KCWI observations of the Leo~P \hii{} region and the \tlusty{} model fit to observations of LP26 does provide one data point supporting the use of \tlusty{} ionizing spectra at extremely low $Z$.

To predict the ionizing photon production by massive stellar populations using SPS codes requires not only stellar spectral templates, but also stellar evolution models that predict a star's \teff{}, \logg{}, and \lstar{} given its age and $Z$.
We have found excellent agreement between the ionizing spectrum predicted by a theoretical \tlusty{} model of similar \teff{} to LP26 and that required to power the observed emission from star's surrounding \hii{} region.
However, this comparison does not directly test the predictions of stellar evolution models at low $Z$, which may be strongly affected by assumptions about various physical processes, including mass loss, mixing, and rotation. 
Previous observational evidence suggests that fast rotation may be more common at lower $Z$ \citep{ramachandran19, telford21}, implying longer stellar lifetimes and a different relationship between \teff{}, \lstar, and age, which together boost lifetime ionizing photon production by fast-rotating stars \citep[e.g.,][]{topping15, murphy21}. 
Thus, to fully validate the SPS-based predictions of ionizing photon production by XMP galaxies will require additional observational constraints on changes in stellar evolutionary pathways at extremely low $Z$.

\begin{figure*}[!htp]
\begin{centering}
  \includegraphics[width=\linewidth]{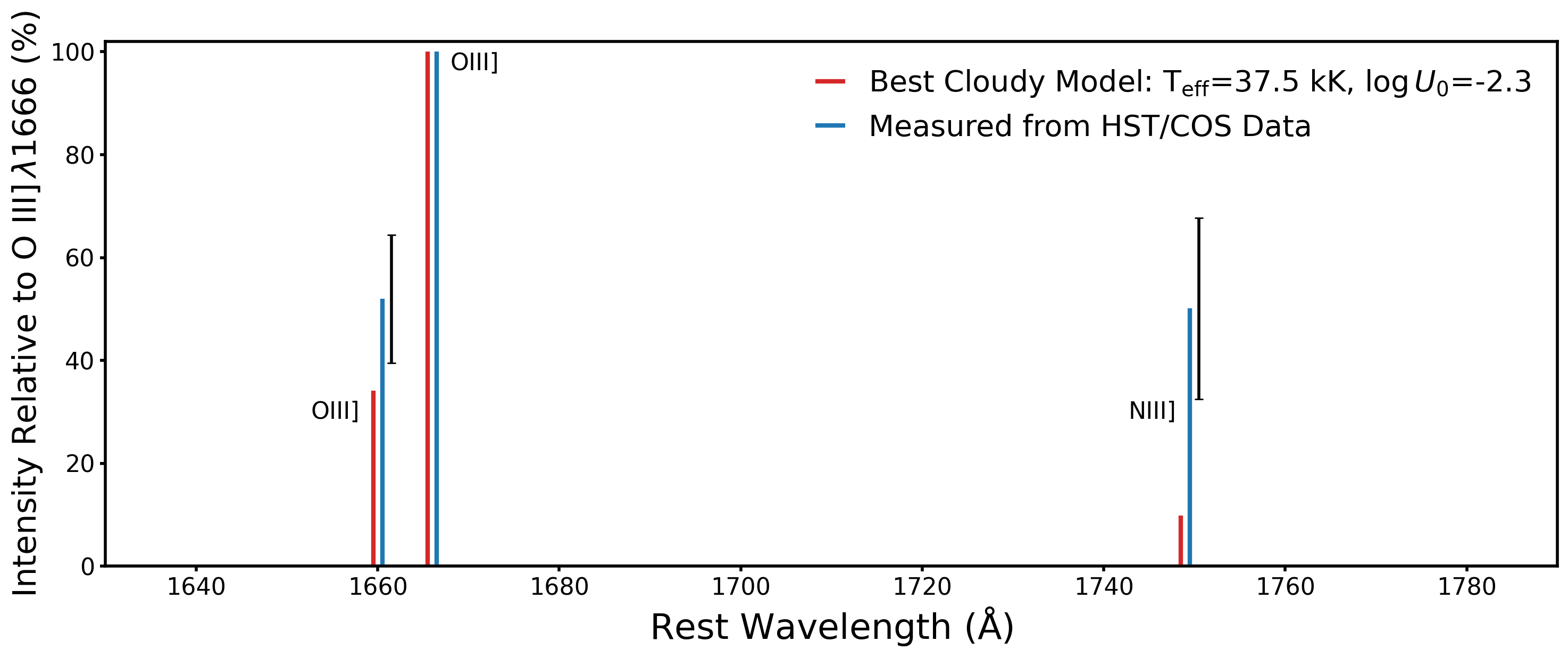}
\caption{\textbf{Photoionization models do not reproduce the unusual FUV N\,\textsc{iii}] emission observed in Leo~P.} 
Vertical red lines represent predicted FUV emission line intensities, relative to the strongest O\,\textsc{iii}]\,$\lambda$1666 line, from the \cloudy{} model fit to the optical KCWI emission, while blue lines show the intensity ratios measured from the HST/COS data. 
Black error bars show the measurement uncertainties on the observations (i.e., not accounting for uncertainty in the extinction correction).
\label{fig:cloudy_fuv}}
\end{centering}
\end{figure*}

\subsection{Puzzles in Leo~P's FUV Nebular Emission\label{sec:fuv_puzzles}}

In Section~\ref{sec:optical_fluxes}, we reported a strong detection of the N\,\textsc{iii}]\,$\lambda\lambda$1748,\,1749 emission lines in the HST/COS spectrum of Leo~P's \hii{} region.
This emission line is very rarely detected in FUV observations of star-forming galaxies, and has rarely been studied in detail. 
The observed emission has two characteristics that are challenging to explain.
First, we found that the wavelength spacing of the two emission features uniquely identifies which two of several possible N\,\textsc{iii}] transitions we are observing. 
We also found that the observed O\,\textsc{iii}] and N\,\textsc{iii}] lines have substantially different velocities, in the sense that the N\,\textsc{iii}] emission is blueshifted by about 57\,km\,s$^{-1}$ (Section~\ref{sec:fuv_fluxes}). 
This velocity offset is comparable to the FWHM of the emission lines (86\,km\,s$^{-1}$), too large to be attributed to wavelength calibration uncertainties.
Yet, phenomena like outflows, stellar winds, and shocks are all associated with higher velocities, from 100s to 1000s of km\,s$^{-1}$. 
It is not obvious what mechanism is driving the velocity offset of the FUV N\,\textsc{iii}] emission in Leo~P.

The second odd property of the N\,\textsc{iii}]\,$\lambda$1750 emission is its weak intensity relative to the predictions of \cloudy{} models that are matched to the measured \otemp{} and N abundance of the Leo~P \hii{} region.
Figure~\ref{fig:cloudy_fuv} presents a comparison between the observed (blue lines) and predicted (red lines) FUV emission line intensities, normalized to the O\,\textsc{iii}]\,$\lambda$1666 intensity (Table~\ref{tab:obs_lines}).
As in Figure~\ref{fig:cloudy_optical}, the vertical lines representing each line's intensity are plotted at its rest wavelength, with a small offset for clarity, and black error bars show the measurement uncertainties on the observed intensities (not accounting for uncertainty in the dust correction). 
We only show the best-fit \cloudy{} model to the emission line intensities measured from the KCWI data, but the following discussion holds across all combinations of $\log U_0$ and stellar \teff{} that we tested in the photoionization models. 
While the O\,\textsc{iii}]\,$\lambda$1660/O\,\textsc{iii}]\,$\lambda$1666 ratio predicted by \cloudy{} does not quite agree within the measurement uncertainties, that modest mismatch is plausibly explained by uncertainties in the atomic data for the collisionally excited O\,\textsc{iii}].
The observed N\,\textsc{iii}]\,$\lambda$1750 emission, however, is far stronger than predicted by \textit{any} of the \cloudy{} models that we tested.

We therefore consider alternative scenarios to the assumption that the observed N\,\textsc{iii}] emission is nebular in nature. 
P-Cygni wind profiles of N\,\textsc{iii}\,$\lambda$1750 are sometimes seen in supergiant or Wolf-Rayet stars with very high \mdot{} (e.g., \citealt{crowther02}), but we see no evidence of that characteristic blueshifted absorption combined with redshifted emission.
Redshifted wind emission also cannot account for the blueshifted N\,\textsc{iii}] emission relative to both the O\,\textsc{iii}] lines and the systemic velocity of Leo~P (Table~\ref{tab:properties}). 
The wind-sensitive N\,\textsc{v}\,$\lambda$1240 and C\,\textsc{iv}\,$\lambda$1550 lines in the FUV spectrum of LP26 are extremely weak \citep{telford21}, so this star must be driving a low-\mdot{} wind.
Furthermore, emission from a high-velocity wind (reaching $\sim$\,1000s of km\,s$^{-1}$) would be quite broad, and thus cannot explain the observed narrow N\,\textsc{iii}]\,$\lambda\lambda$1748,\,1749 emission. 

Photospheric absorption in N\,\textsc{iii}\,$\lambda\lambda$1748,\,1752 is common in main-sequence O stars \citep[e.g.,][]{walborn85}, so narrow absorption lines could be blended with the nebular emission to produce apparently lower emission line fluxes.
But this would actually make the mismatch between observations and \cloudy{} worse, as the true emission in Leo~P would be even more dramatically under-predicted by the photoionization models.
Moreover, photospheric absorption in LP26 is unlikely to impact the observed N\,\textsc{iii}] emission for two reasons. 
First, \citet{telford21} did not detect photospheric N\,\textsc{iii}\,$\lambda\lambda$1183,\,1184 absorption in the bluer, higher-SNR part of the HST/COS spectrum above the continuum noise, and attributed this to both the intrinsic weakness of the metal lines at such low $Z$ and the high measured \vsini{} of the star, which tends to broaden photospheric absorption lines and make them less distinguishable from the continuum level.
Second, the wavelengths of the commonly observed photospheric N\,\textsc{iii} transitions are 1748 and 1752\,\AA{}, so if stellar absorption were significant, we would expect to see a strong absorption feature at 1752\,\AA{}, redward of the observed emission. 
Inspection of Figure~\ref{fig:fuv_spectrum} reveals no evidence of such absorption.

We have no physical scenario to propose that can explain the detection of this anomalous FUV N\,\textsc{iii}] nebular emission in Leo~P, its modest blueshift with respect to the O\,\textsc{iii}] lines and systemic \hi{} velocity, and its strong intensity relative to the predictions of photoionization models fit to the optical nebular emission. 
Rather, we highlight these findings in the first FUV spectrum of an extremely metal-poor \hii{} region powered by an apparently single star as motivation for future study.


\section{Conclusions\label{sec:conclusions}}

In this paper, we presented new Keck/KCWI optical IFU spectroscopy (Section~\ref{sec:kcwi_data}, Figures~\ref{fig:kcwi_cubes} and \ref{fig:extracted_spectra}) of the only \hii{} region in the 3\%\,\zsun{} galaxy Leo~P, which is powered by a single O star, LP26.
We used the observed emission line fluxes to measure key nebular properties and to constrain the normalization and shape of the ionizing stellar spectrum powering that emission.
We then ran \cloudy{} photoionization models to test the ability of theoretical \tlusty{} spectra at extremely low-$Z$ to reproduce emission lines produced by ions across a wide range of ionization energies.
Finally, we measured nebular emission line fluxes in the HST/COS FUV spectrum of the Leo~P \hii{} region and compared to the predictions of the \cloudy{} models.
Our conclusions are:

\begin{enumerate}[noitemsep, topsep=0pt]

\item We modeled the optical and FUV emission lines from the Leo~P \hii{} region as Gaussians to measure their fluxes (Section~\ref{sec:emlines}, Figures~\ref{fig:optical_gaussian_fits} and \ref{fig:fuv_spectrum}, Table~\ref{tab:obs_lines}). 
From the observed Balmer line ratios and strength of the [O\,\textsc{iii}]\,$\lambda$4363 auroral line relative to [O\,\textsc{iii}]\,$\lambda\lambda$4959,\,5007, we determined the nebular \ebv{} and \otemp{}, confirming the hot, metal-poor, and low-dust nature of the \hii{} region (Section~\ref{sec:neb_conditions}, Table~\ref{tab:results}). 

\item We calculated the H-ionizing photon production rate of LP26 required to power the observed, dust-corrected H$\beta$ intensity: $Q(\mathrm{H}) = (3.74 \pm 0.67) \times 10^{48}$\,s$^{-1}$ (Section~\ref{sec:qion}).
We compared to the prediction of the \tlusty{}/\textsc{parsec} model SED fit to the FUV spectrum and optical/NIR HST photometry of LP26 \citep{telford21}, and found excellent agreement: our measurement is within the observational uncertainties of, and just 13\% lower than, the model-predicted \qh{}.

\item To measure the hardness of the ionizing spectrum, we similarly calculated LP26's production rate of more energetic, He-ionizing photons from the observed He\,\textsc{i}\,$\lambda$4471 intensity.
The \qhei{}/\qh{}\,$=$\,$0.12\pm0.03$ that we measure is lower than the 0.18 predicted by the \tlusty{} model fit to the star's SED.
This suggests that LP26 may be slightly cooler than the best-fit \teff{} of 37.5\,kK, but still within one step in the coarsely sampled \textsc{ostar2002} model grid (Section~\ref{sec:q_ratios}, Figure~\ref{fig:qion_vs_teff}).

\item We ran \cloudy{} photoionization models tailored to the known properties of the Leo~P \hii{} region with \tlusty{} spectra of varying \teff{} as the ionizing source. 
Comparing the models to the strengths of all emission lines detected in the KCWI data, which require ionization energies from 10.4\,eV (S$^{+}$) to 40.7\,eV (Ar$^{3+}$), we find that the 37.5\,kK model provides the best overall match to the data (Section~\ref{sec:cloudy}, Figure~\ref{fig:cloudy_optical}).
Together with the \qhei{}/\qh{} constraint, our results suggest that LP26's \teff{} is between 35 and 37.5\,kK, and that the 2\%\,\zsun{}, wind-free \tlusty{} model stellar spectra agree remarkably well with the nebular emission observed from Leo~P's \hii{} region.
The HST/COS FUV spectrum of LP26 shows that its stellar wind is very weak \citep{telford21}, potentially explaining the consistency with models that neglect the impacts of mass loss (Section~\ref{sec:implications}). 

\item We report a rare detection of N\,\textsc{iii}]\,$\lambda$1749 emission in the  FUV spectrum of the Leo~P \hii{} region. 
The observed strength of the N\,\textsc{iii}] emission relative to FUV O\,\textsc{iii}]\,$\lambda\lambda$1660,\,1666 lines is stronger than predicted by any of the \cloudy{} models we tested, and the N\,\textsc{iii}] lines are blueshifted by 57\,km\,s$^{-1}$ relative to the O\,\textsc{iii}] lines.
The physical driver of the N\,\textsc{iii}] emission remains unclear and warrants future study
(Sections~\ref{sec:fuv_fluxes} and \ref{sec:fuv_puzzles}, Figure~\ref{fig:cloudy_fuv}).

\end{enumerate} 

This first analysis of a single-star \hii{} region in an XMP galaxy has provided empirical support for the use of \tlusty{} spectra to model massive stellar populations at 2-3\% \zsun{}. 
While previous work at higher $Z$ has found that \tlusty{} models that do not treat the impacts of expanding stellar atmospheres produce ionizing spectra that are too soft \citep[e.g.,][]{zastrow13, martins21}, we find good agreement between the \tlusty{} models and our KCWI observations of Leo~P's \hii{} reigion.
We suggest that stellar winds have negligible effects on the ionizing spectra of typical (i.e., moderate \mstar{} and \teff{}) O stars in the XMP regime, but confirmation awaits future studies of other single-star \hii{} regions more metal-poor than the 20\%\,\zsun{} SMC.


\acknowledgements

This work was supported by a NASA Keck PI Data Award, administered by the NASA Exoplanet Science Institute. Data presented herein were obtained at the W. M. Keck Observatory from telescope time allocated to the National Aeronautics and Space Administration through the agency's scientific partnership with the California Institute of Technology and the University of California. The Observatory was made possible by the generous financial support of the W. M. Keck Foundation.
The authors wish to recognize and acknowledge the very significant cultural role and reverence that the summit of Maunakea has always had within the indigenous Hawaiian community. We are most fortunate to have the opportunity to conduct observations from this mountain.

This work was supported by the Space Telescope Science
Institute through GO-15967. 
This work was performed in part at Aspen Center for Physics, which is supported by National Science Foundation grant PHY-1607611, and was partially supported by a grant from the Simons Foundation.
This research has extensively used NASA's Astrophysics Data System, adstex\footnote{\url{https://github.com/yymao/adstex}}, and the arXiv preprint server. 

\facilities{Keck:II (KCWI), HST (COS)}

\software{Astropy \citep{astropy, astropy2, astropy3}, \cloudy{} \citep[][other appropriate citations]{ferland98}, CWITools \citep{osullivan20}, HDF5 \citep{hdf5},  KCWI DRP \citep{morrissey18}, iPython \citep{ipython}, Matplotlib \citep{matplotlib}, NumPy \citep{numpy2}, \pycloudy{} \citep{morisset13}, \pyneb{} \citep{luridiana12, luridiana15}, SAOImageDS9 \citep{ds9}, SciPy \citep{scipy2}}


\begin{thebibliography}{}
\expandafter\ifx\csname natexlab\endcsname\relax\def\natexlab#1{#1}\fi
\providecommand{\url}[1]{\href{#1}{#1}}
\providecommand{\dodoi}[1]{doi:~\href{http://doi.org/#1}{\nolinkurl{#1}}}
\providecommand{\doeprint}[1]{\href{http://ascl.net/#1}{\nolinkurl{http://ascl.net/#1}}}
\providecommand{\doarXiv}[1]{\href{https://arxiv.org/abs/#1}{\nolinkurl{https://arxiv.org/abs/#1}}}

\bibitem[{{Aggarwal} \& {Keenan}(1999)}]{aggarwal99}
{Aggarwal}, K.~M., \& {Keenan}, F.~P. 1999, \apjs, 123, 311,
  \dodoi{10.1086/313232}

\bibitem[{{Ali} {et~al.}(1991){Ali}, {Blum}, {Bumgardner}, {Cranmer},
  {Ferland}, {Haefner}, \& {Tiede}}]{ali91}
{Ali}, B., {Blum}, R.~D., {Bumgardner}, T.~E., {et~al.} 1991, \pasp, 103, 1182,
  \dodoi{10.1086/132938}

\bibitem[{{Asplund} {et~al.}(2009){Asplund}, {Grevesse}, {Sauval}, \&
  {Scott}}]{asplund09}
{Asplund}, M., {Grevesse}, N., {Sauval}, A.~J., \& {Scott}, P. 2009, \araa, 47,
  481, \dodoi{10.1146/annurev.astro.46.060407.145222}

\bibitem[{{Astropy Collaboration} {et~al.}(2013){Astropy Collaboration},
  {Robitaille}, {Tollerud}, {Greenfield}, {Droettboom}, {Bray}, {Aldcroft},
  {Davis}, {Ginsburg}, {Price-Whelan}, {Kerzendorf}, {Conley}, {Crighton},
  {Barbary}, {Muna}, {Ferguson}, {Grollier}, {Parikh}, {Nair}, {Unther},
  {Deil}, {Woillez}, {Conseil}, {Kramer}, {Turner}, {Singer}, {Fox}, {Weaver},
  {Zabalza}, {Edwards}, {Azalee Bostroem}, {Burke}, {Casey}, {Crawford},
  {Dencheva}, {Ely}, {Jenness}, {Labrie}, {Lim}, {Pierfederici}, {Pontzen},
  {Ptak}, {Refsdal}, {Servillat}, \& {Streicher}}]{astropy}
{Astropy Collaboration}, {Robitaille}, T.~P., {Tollerud}, E.~J., {et~al.} 2013,
  \aap, 558, A33, \dodoi{10.1051/0004-6361/201322068}

\bibitem[{{Astropy Collaboration} {et~al.}(2018){Astropy Collaboration},
  {Price-Whelan}, {Sip{\H{o}}cz}, {G{\"u}nther}, {Lim}, {Crawford}, {Conseil},
  {Shupe}, {Craig}, {Dencheva}, {Ginsburg}, {VanderPlas}, {Bradley},
  {P{\'e}rez-Su{\'a}rez}, {de Val-Borro}, {Aldcroft}, {Cruz}, {Robitaille},
  {Tollerud}, {Ardelean}, {Babej}, {Bach}, {Bachetti}, {Bakanov}, {Bamford},
  {Barentsen}, {Barmby}, {Baumbach}, {Berry}, {Biscani}, {Boquien}, {Bostroem},
  {Bouma}, {Brammer}, {Bray}, {Breytenbach}, {Buddelmeijer}, {Burke},
  {Calderone}, {Cano Rodr{\'\i}guez}, {Cara}, {Cardoso}, {Cheedella}, {Copin},
  {Corrales}, {Crichton}, {D'Avella}, {Deil}, {Depagne}, {Dietrich}, {Donath},
  {Droettboom}, {Earl}, {Erben}, {Fabbro}, {Ferreira}, {Finethy}, {Fox},
  {Garrison}, {Gibbons}, {Goldstein}, {Gommers}, {Greco}, {Greenfield},
  {Groener}, {Grollier}, {Hagen}, {Hirst}, {Homeier}, {Horton}, {Hosseinzadeh},
  {Hu}, {Hunkeler}, {Ivezi{\'c}}, {Jain}, {Jenness}, {Kanarek}, {Kendrew},
  {Kern}, {Kerzendorf}, {Khvalko}, {King}, {Kirkby}, {Kulkarni}, {Kumar},
  {Lee}, {Lenz}, {Littlefair}, {Ma}, {Macleod}, {Mastropietro}, {McCully},
  {Montagnac}, {Morris}, {Mueller}, {Mumford}, {Muna}, {Murphy}, {Nelson},
  {Nguyen}, {Ninan}, {N{\"o}the}, {Ogaz}, {Oh}, {Parejko}, {Parley}, {Pascual},
  {Patil}, {Patil}, {Plunkett}, {Prochaska}, {Rastogi}, {Reddy Janga},
  {Sabater}, {Sakurikar}, {Seifert}, {Sherbert}, {Sherwood-Taylor}, {Shih},
  {Sick}, {Silbiger}, {Singanamalla}, {Singer}, {Sladen}, {Sooley},
  {Sornarajah}, {Streicher}, {Teuben}, {Thomas}, {Tremblay}, {Turner},
  {Terr{\'o}n}, {van Kerkwijk}, {de la Vega}, {Watkins}, {Weaver}, {Whitmore},
  {Woillez}, {Zabalza}, \& {Astropy Contributors}}]{astropy2}
{Astropy Collaboration}, {Price-Whelan}, A.~M., {Sip{\H{o}}cz}, B.~M., {et~al.}
  2018, \aj, 156, 123, \dodoi{10.3847/1538-3881/aabc4f}

\bibitem[{{Astropy Collaboration} {et~al.}(2022){Astropy Collaboration},
  {Price-Whelan}, {Lim}, {Earl}, {Starkman}, {Bradley}, {Shupe}, {Patil},
  {Corrales}, {Brasseur}, {N{\"o}the}, {Donath}, {Tollerud}, {Morris},
  {Ginsburg}, {Vaher}, {Weaver}, {Tocknell}, {Jamieson}, {van Kerkwijk},
  {Robitaille}, {Merry}, {Bachetti}, {G{\"u}nther}, {Aldcroft},
  {Alvarado-Montes}, {Archibald}, {B{\'o}di}, {Bapat}, {Barentsen},
  {Baz{\'a}n}, {Biswas}, {Boquien}, {Burke}, {Cara}, {Cara}, {Conroy},
  {Conseil}, {Craig}, {Cross}, {Cruz}, {D'Eugenio}, {Dencheva}, {Devillepoix},
  {Dietrich}, {Eigenbrot}, {Erben}, {Ferreira}, {Foreman-Mackey}, {Fox},
  {Freij}, {Garg}, {Geda}, {Glattly}, {Gondhalekar}, {Gordon}, {Grant},
  {Greenfield}, {Groener}, {Guest}, {Gurovich}, {Handberg}, {Hart},
  {Hatfield-Dodds}, {Homeier}, {Hosseinzadeh}, {Jenness}, {Jones}, {Joseph},
  {Kalmbach}, {Karamehmetoglu}, {Ka{\l}uszy{\'n}ski}, {Kelley}, {Kern},
  {Kerzendorf}, {Koch}, {Kulumani}, {Lee}, {Ly}, {Ma}, {MacBride}, {Maljaars},
  {Muna}, {Murphy}, {Norman}, {O'Steen}, {Oman}, {Pacifici}, {Pascual},
  {Pascual-Granado}, {Patil}, {Perren}, {Pickering}, {Rastogi}, {Roulston},
  {Ryan}, {Rykoff}, {Sabater}, {Sakurikar}, {Salgado}, {Sanghi}, {Saunders},
  {Savchenko}, {Schwardt}, {Seifert-Eckert}, {Shih}, {Jain}, {Shukla}, {Sick},
  {Simpson}, {Singanamalla}, {Singer}, {Singhal}, {Sinha}, {Sip{\H{o}}cz},
  {Spitler}, {Stansby}, {Streicher}, {{\v{S}}umak}, {Swinbank}, {Taranu},
  {Tewary}, {Tremblay}, {Val-Borro}, {Van Kooten}, {Vasovi{\'c}}, {Verma}, {de
  Miranda Cardoso}, {Williams}, {Wilson}, {Winkel}, {Wood-Vasey}, {Xue},
  {Yoachim}, {Zhang}, {Zonca}, \& {Astropy Project Contributors}}]{astropy3}
{Astropy Collaboration}, {Price-Whelan}, A.~M., {Lim}, P.~L., {et~al.} 2022,
  \apj, 935, 167, \dodoi{10.3847/1538-4357/ac7c74}

\bibitem[{{Aver} {et~al.}(2022){Aver}, {Berg}, {Hirschauer}, {Olive}, {Pogge},
  {Rogers}, {Salzer}, \& {Skillman}}]{aver22}
{Aver}, E., {Berg}, D.~A., {Hirschauer}, A.~S., {et~al.} 2022, \mnras, 510,
  373, \dodoi{10.1093/mnras/stab3226}

\bibitem[{{Becker} {et~al.}(2001){Becker}, {Fan}, {White}, {Strauss},
  {Narayanan}, {Lupton}, {Gunn}, {Annis}, {Bahcall}, {Brinkmann}, {Connolly},
  {Csabai}, {Czarapata}, {Doi}, {Heckman}, {Hennessy}, {Ivezi{\'c}}, {Knapp},
  {Lamb}, {McKay}, {Munn}, {Nash}, {Nichol}, {Pier}, {Richards}, {Schneider},
  {Stoughton}, {Szalay}, {Thakar}, \& {York}}]{becker01}
{Becker}, R.~H., {Fan}, X., {White}, R.~L., {et~al.} 2001, \aj, 122, 2850,
  \dodoi{10.1086/324231}

\bibitem[{{Berg} {et~al.}(2022){Berg}, {James}, {King}, {McDonald}, {Chen},
  {Chisholm}, {Heckman}, {Martin}, {Stark}, {Aloisi}, {Amor{\'\i}n},
  {Arellano-C{\'o}rdova}, {Bayliss}, {Bordoloi}, {Brinchmann}, {Charlot},
  {Chevallard}, {Clark}, {Erb}, {Feltre}, {Gronke}, {Hayes}, {Henry},
  {Hernandez}, {Jaskot}, {Jones}, {Kewley}, {Kumari}, {Leitherer}, {Llerena},
  {Maseda}, {Mingozzi}, {Nanayakkara}, {Ouchi}, {Plat}, {Pogge},
  {Ravindranath}, {Rigby}, {Sanders}, {Scarlata}, {Senchyna}, {Skillman},
  {Steidel}, {Strom}, {Sugahara}, {Wilkins}, {Wofford}, {Xu}, \& {Classy
  Team}}]{berg22}
{Berg}, D.~A., {James}, B.~L., {King}, T., {et~al.} 2022, \apjs, 261, 31,
  \dodoi{10.3847/1538-4365/ac6c03}

\bibitem[{{Bernstein-Cooper} {et~al.}(2014){Bernstein-Cooper}, {Cannon},
  {Elson}, {Warren}, {Chengular}, {Skillman}, {Adams}, {Bolatto}, {Giovanelli},
  {Haynes}, {McQuinn}, {Pardy}, {Rhode}, \& {Salzer}}]{bernstein-cooper14}
{Bernstein-Cooper}, E.~Z., {Cannon}, J.~M., {Elson}, E.~C., {et~al.} 2014, \aj,
  148, 35, \dodoi{10.1088/0004-6256/148/2/35}

\bibitem[{{Bj{\"o}rklund} {et~al.}(2021){Bj{\"o}rklund}, {Sundqvist}, {Puls},
  \& {Najarro}}]{bjorklund21}
{Bj{\"o}rklund}, R., {Sundqvist}, J.~O., {Puls}, J., \& {Najarro}, F. 2021,
  \aap, 648, A36, \dodoi{10.1051/0004-6361/202038384}

\bibitem[{{Bressan} {et~al.}(2012){Bressan}, {Marigo}, {Girardi}, {Salasnich},
  {Dal Cero}, {Rubele}, \& {Nanni}}]{bressan12}
{Bressan}, A., {Marigo}, P., {Girardi}, L., {et~al.} 2012, \mnras, 427, 127,
  \dodoi{10.1111/j.1365-2966.2012.21948.x}

\bibitem[{{Byler} {et~al.}(2017){Byler}, {Dalcanton}, {Conroy}, \&
  {Johnson}}]{byler17}
{Byler}, N., {Dalcanton}, J.~J., {Conroy}, C., \& {Johnson}, B.~D. 2017, \apj,
  840, 44, \dodoi{10.3847/1538-4357/aa6c66}

\bibitem[{{Cardelli} {et~al.}(1989){Cardelli}, {Clayton}, \&
  {Mathis}}]{cardelli89}
{Cardelli}, J.~A., {Clayton}, G.~C., \& {Mathis}, J.~S. 1989, \apj, 345, 245,
  \dodoi{10.1086/167900}

\bibitem[{{Chisholm} {et~al.}(2019){Chisholm}, {Rigby}, {Bayliss}, {Berg},
  {Dahle}, {Gladders}, \& {Sharon}}]{chisholm19}
{Chisholm}, J., {Rigby}, J.~R., {Bayliss}, M., {et~al.} 2019, \apj, 882, 182,
  \dodoi{10.3847/1538-4357/ab3104}

\bibitem[{{Chisholm} {et~al.}(2022){Chisholm}, {Saldana-Lopez}, {Flury},
  {Schaerer}, {Jaskot}, {Amor{\'\i}n}, {Atek}, {Finkelstein}, {Fleming},
  {Ferguson}, {Fern{\'a}ndez}, {Giavalisco}, {Hayes}, {Heckman}, {Henry}, {Ji},
  {Marques-Chaves}, {Mauerhofer}, {McCandliss}, {Oey}, {{\"O}stlin},
  {Rutkowski}, {Scarlata}, {Thuan}, {Trebitsch}, {Wang}, {Worseck}, \&
  {Xu}}]{chisholm22}
{Chisholm}, J., {Saldana-Lopez}, A., {Flury}, S., {et~al.} 2022, \mnras,
  \dodoi{10.1093/mnras/stac2874}

\bibitem[{{Conroy}(2013)}]{conroy13}
{Conroy}, C. 2013, \araa, 51, 393, \dodoi{10.1146/annurev-astro-082812-141017}

\bibitem[{{Crowther} {et~al.}(2002){Crowther}, {Hillier}, {Evans}, {Fullerton},
  {De Marco}, \& {Willis}}]{crowther02}
{Crowther}, P.~A., {Hillier}, D.~J., {Evans}, C.~J., {et~al.} 2002, \apj, 579,
  774, \dodoi{10.1086/342877}

\bibitem[{{Dayal} \& {Ferrara}(2018)}]{dayal18}
{Dayal}, P., \& {Ferrara}, A. 2018, \physrep, 780, 1,
  \dodoi{10.1016/j.physrep.2018.10.002}

\bibitem[{{Dors} \& {Copetti}(2003)}]{dors03}
{Dors}, O.~L., J., \& {Copetti}, M.~V.~F. 2003, \aap, 404, 969,
  \dodoi{10.1051/0004-6361:20030636}

\bibitem[{{Dufour}(1984)}]{dufour84}
{Dufour}, R.~J. 1984, in Structure and Evolution of the Magellanic Clouds, ed.
  S.~{van den Bergh} \& K.~S.~D. {de Boer}, Vol. 108, 353--361,
  \dodoi{10.1017/S0074180900040481}

\bibitem[{{Endsley} {et~al.}(2021){Endsley}, {Stark}, {Chevallard}, \&
  {Charlot}}]{endsley21}
{Endsley}, R., {Stark}, D.~P., {Chevallard}, J., \& {Charlot}, S. 2021, \mnras,
  500, 5229, \dodoi{10.1093/mnras/staa3370}

\bibitem[{{Endsley} {et~al.}(2022){Endsley}, {Stark}, {Whitler}, {Topping},
  {Chen}, {Plat}, {Chisholm}, \& {Charlot}}]{endsley22}
{Endsley}, R., {Stark}, D.~P., {Whitler}, L., {et~al.} 2022, arXiv e-prints,
  arXiv:2208.14999.
\newblock \doarXiv{2208.14999}

\bibitem[{{Erb}(2015)}]{erb15}
{Erb}, D.~K. 2015, \nat, 523, 169, \dodoi{10.1038/nature14454}

\bibitem[{{Evans} {et~al.}(2019){Evans}, {Castro}, {Gonzalez}, {Garcia},
  {Bastian}, {Cioni}, {Clark}, {Davies}, {Ferguson}, {Kamann}, {Lennon},
  {Patrick}, {Vink}, \& {Weisz}}]{evans19}
{Evans}, C.~J., {Castro}, N., {Gonzalez}, O.~A., {et~al.} 2019, \aap, 622,
  A129, \dodoi{10.1051/0004-6361/201834145}

\bibitem[{{Ferland} {et~al.}(1998){Ferland}, {Korista}, {Verner}, {Ferguson},
  {Kingdon}, \& {Verner}}]{ferland98}
{Ferland}, G.~J., {Korista}, K.~T., {Verner}, D.~A., {et~al.} 1998, \pasp, 110,
  761, \dodoi{10.1086/316190}

\bibitem[{{Ferland} {et~al.}(2017){Ferland}, {Chatzikos}, {Guzm{\'a}n},
  {Lykins}, {van Hoof}, {Williams}, {Abel}, {Badnell}, {Keenan}, {Porter}, \&
  {Stancil}}]{ferland17}
{Ferland}, G.~J., {Chatzikos}, M., {Guzm{\'a}n}, F., {et~al.} 2017, \rmxaa, 53,
  385.
\newblock \doarXiv{1705.10877}

\bibitem[{{Finkelstein} {et~al.}(2019){Finkelstein}, {D'Aloisio},
  {Paardekooper}, {Ryan}, {Behroozi}, {Finlator}, {Livermore}, {Upton
  Sanderbeck}, {Dalla Vecchia}, \& {Khochfar}}]{finkelstein19}
{Finkelstein}, S.~L., {D'Aloisio}, A., {Paardekooper}, J.-P., {et~al.} 2019,
  \apj, 879, 36, \dodoi{10.3847/1538-4357/ab1ea8}

\bibitem[{{Flury} {et~al.}(2022){Flury}, {Jaskot}, {Ferguson}, {Worseck},
  {Makan}, {Chisholm}, {Saldana-Lopez}, {Schaerer}, {McCandliss}, {Wang},
  {Ford}, {Heckman}, {Ji}, {Giavalisco}, {Amorin}, {Atek}, {Blaizot},
  {Borthakur}, {Carr}, {Castellano}, {Cristiani}, {De Barros}, {Dickinson},
  {Finkelstein}, {Fleming}, {Fontanot}, {Garel}, {Grazian}, {Hayes}, {Henry},
  {Mauerhofer}, {Micheva}, {Oey}, {Ostlin}, {Papovich}, {Pentericci},
  {Ravindranath}, {Rosdahl}, {Rutkowski}, {Santini}, {Scarlata}, {Teplitz},
  {Thuan}, {Trebitsch}, {Vanzella}, {Verhamme}, \& {Xu}}]{flury22}
{Flury}, S.~R., {Jaskot}, A.~E., {Ferguson}, H.~C., {et~al.} 2022, \apjs, 260,
  1, \dodoi{10.3847/1538-4365/ac5331}

\bibitem[{{Giovanelli} {et~al.}(2005){Giovanelli}, {Haynes}, {Kent},
  {Perillat}, {Saintonge}, {Brosch}, {Catinella}, {Hoffman}, {Stierwalt},
  {Spekkens}, {Lerner}, {Masters}, {Momjian}, {Rosenberg}, {Springob},
  {Boselli}, {Charmandaris}, {Darling}, {Davies}, {Garcia Lambas}, {Gavazzi},
  {Giovanardi}, {Hardy}, {Hunt}, {Iovino}, {Karachentsev}, {Karachentseva},
  {Koopmann}, {Marinoni}, {Minchin}, {Muller}, {Putman}, {Pantoja}, {Salzer},
  {Scodeggio}, {Skillman}, {Solanes}, {Valotto}, {van Driel}, \& {van
  Zee}}]{giovanelli05}
{Giovanelli}, R., {Haynes}, M.~P., {Kent}, B.~R., {et~al.} 2005, \aj, 130,
  2598, \dodoi{10.1086/497431}

\bibitem[{{Giovanelli} {et~al.}(2013){Giovanelli}, {Haynes}, {Adams}, {Cannon},
  {Rhode}, {Salzer}, {Skillman}, {Bernstein-Cooper}, \&
  {McQuinn}}]{giovanelli13}
{Giovanelli}, R., {Haynes}, M.~P., {Adams}, E. A.~K., {et~al.} 2013, \aj, 146,
  15, \dodoi{10.1088/0004-6256/146/1/15}

\bibitem[{{Green} {et~al.}(2015){Green}, {Schlafly}, {Finkbeiner}, {Rix},
  {Martin}, {Burgett}, {Draper}, {Flewelling}, {Hodapp}, {Kaiser}, {Kudritzki},
  {Magnier}, {Metcalfe}, {Price}, {Tonry}, \& {Wainscoat}}]{green15}
{Green}, G.~M., {Schlafly}, E.~F., {Finkbeiner}, D.~P., {et~al.} 2015, \apj,
  810, 25, \dodoi{10.1088/0004-637X/810/1/25}

\bibitem[{{Grevesse} {et~al.}(2010){Grevesse}, {Asplund}, {Sauval}, \&
  {Scott}}]{grevesse10}
{Grevesse}, N., {Asplund}, M., {Sauval}, A.~J., \& {Scott}, P. 2010, \apss,
  328, 179, \dodoi{10.1007/s10509-010-0288-z}

\bibitem[{{Grevesse} \& {Sauval}(1998)}]{grevesse98}
{Grevesse}, N., \& {Sauval}, A.~J. 1998, \ssr, 85, 161,
  \dodoi{10.1023/A:1005161325181}

\bibitem[{{Harris} {et~al.}(2020){Harris}, {Millman}, {van der Walt},
  {Gommers}, {Virtanen}, {Cournapeau}, {Wieser}, {Taylor}, {Berg}, {Smith},
  {Kern}, {Picus}, {Hoyer}, {van Kerkwijk}, {Brett}, {Haldane}, {del R{\'\i}o},
  {Wiebe}, {Peterson}, {G{\'e}rard-Marchant}, {Sheppard}, {Reddy}, {Weckesser},
  {Abbasi}, {Gohlke}, \& {Oliphant}}]{numpy2}
{Harris}, C.~R., {Millman}, K.~J., {van der Walt}, S.~J., {et~al.} 2020, \nat,
  585, 357, \dodoi{10.1038/s41586-020-2649-2}

\bibitem[{{Haynes} {et~al.}(2011){Haynes}, {Giovanelli}, {Martin}, {Hess},
  {Saintonge}, {Adams}, {Hallenbeck}, {Hoffman}, {Huang}, {Kent}, {Koopmann},
  {Papastergis}, {Stierwalt}, {Balonek}, {Craig}, {Higdon}, {Kornreich},
  {Miller}, {O'Donoghue}, {Olowin}, {Rosenberg}, {Spekkens}, {Troischt}, \&
  {Wilcots}}]{haynes11}
{Haynes}, M.~P., {Giovanelli}, R., {Martin}, A.~M., {et~al.} 2011, \aj, 142,
  170, \dodoi{10.1088/0004-6256/142/5/170}

\bibitem[{{Hill} {et~al.}(2014){Hill}, {Benjamin}, {Haffner}, {Gostisha}, \&
  {Barger}}]{hill14}
{Hill}, A.~S., {Benjamin}, R.~A., {Haffner}, L.~M., {Gostisha}, M.~C., \&
  {Barger}, K.~A. 2014, \apj, 787, 106, \dodoi{10.1088/0004-637X/787/2/106}

\bibitem[{{Hubeny} \& {Lanz}(1995)}]{hubeny95}
{Hubeny}, I., \& {Lanz}, T. 1995, \apj, 439, 875, \dodoi{10.1086/175226}

\bibitem[{Hunter(2007)}]{matplotlib}
Hunter, J.~D. 2007, Computing In Science \& Engineering, 9, 90

\bibitem[{{Hutchison} {et~al.}(2019){Hutchison}, {Papovich}, {Finkelstein},
  {Dickinson}, {Jung}, {Zitrin}, {Ellis}, {Malhotra}, {Rhoads},
  {Roberts-Borsani}, {Song}, \& {Tilvi}}]{hutchison19}
{Hutchison}, T.~A., {Papovich}, C., {Finkelstein}, S.~L., {et~al.} 2019, \apj,
  879, 70, \dodoi{10.3847/1538-4357/ab22a2}

\bibitem[{{Joye} \& {Mandel}(2003)}]{ds9}
{Joye}, W.~A., \& {Mandel}, E. 2003, in Astronomical Society of the Pacific
  Conference Series, Vol. 295, Astronomical Data Analysis Software and Systems
  XII, ed. H.~E. {Payne}, R.~I. {Jedrzejewski}, \& R.~N. {Hook}, 489

\bibitem[{{Kennicutt} {et~al.}(2000){Kennicutt}, {Bresolin}, {French}, \&
  {Martin}}]{kennicutt00}
{Kennicutt}, Robert~C., J., {Bresolin}, F., {French}, H., \& {Martin}, P. 2000,
  \apj, 537, 589, \dodoi{10.1086/309075}

\bibitem[{{Lanz} \& {Hubeny}(2003)}]{lanz03}
{Lanz}, T., \& {Hubeny}, I. 2003, \apjs, 146, 417, \dodoi{10.1086/374373}

\bibitem[{{Luridiana} {et~al.}(2012){Luridiana}, {Morisset}, \&
  {Shaw}}]{luridiana12}
{Luridiana}, V., {Morisset}, C., \& {Shaw}, R.~A. 2012, IAU Symposium, 283,
  422, \dodoi{10.1017/S1743921312011738}

\bibitem[{{Luridiana} {et~al.}(2015){Luridiana}, {Morisset}, \&
  {Shaw}}]{luridiana15}
---. 2015, \aap, 573, A42, \dodoi{10.1051/0004-6361/201323152}

\bibitem[{{Martins} \& {Palacios}(2021)}]{martins21}
{Martins}, F., \& {Palacios}, A. 2021, \aap, 645, A67,
  \dodoi{10.1051/0004-6361/202039337}

\bibitem[{{Martins} {et~al.}(2005){Martins}, {Schaerer}, {Hillier},
  {Meynadier}, {Heydari-Malayeri}, \& {Walborn}}]{martins05}
{Martins}, F., {Schaerer}, D., {Hillier}, D.~J., {et~al.} 2005, \aap, 441, 735,
  \dodoi{10.1051/0004-6361:20052927}

\bibitem[{{Massey}(2013)}]{massey13}
{Massey}, P. 2013, \nar, 57, 14, \dodoi{10.1016/j.newar.2013.05.002}

\bibitem[{{McQuinn} {et~al.}(2015){McQuinn}, {Skillman}, {Dolphin}, {Cannon},
  {Salzer}, {Rhode}, {Adams}, {Berg}, {Giovanelli}, {Girardi}, \&
  {Haynes}}]{mcquinn15f}
{McQuinn}, K. B.~W., {Skillman}, E.~D., {Dolphin}, A., {et~al.} 2015, \apj,
  812, 158, \dodoi{10.1088/0004-637X/812/2/158}

\bibitem[{{McQuinn} {et~al.}(2020){McQuinn}, {Berg}, {Skillman}, {Adams},
  {Cannon}, {Dolphin}, {Salzer}, {Giovanelli}, {Haynes}, {Hirschauer},
  {Janoweicki}, {Klapkowski}, \& {Rhode}}]{mcquinn20}
{McQuinn}, K. B.~W., {Berg}, D.~A., {Skillman}, E.~D., {et~al.} 2020, \apj,
  891, 181, \dodoi{10.3847/1538-4357/ab7447}

\bibitem[{{Mingozzi} {et~al.}(2022){Mingozzi}, {James}, {Arellano-C{\'o}rdova},
  {Berg}, {Senchyna}, {Chisholm}, {Brinchmann}, {Aloisi}, {Amor{\'\i}n},
  {Charlot}, {Feltre}, {Hayes}, {Heckman}, {Henry}, {Hernandez}, {Kumari},
  {Leitherer}, {Llerena}, {Martin}, {Nanayakkara}, {Ravindranath}, {Skillman},
  {Sugahara}, {Wofford}, \& {Xu}}]{mingozzi22}
{Mingozzi}, M., {James}, B.~L., {Arellano-C{\'o}rdova}, K.~Z., {et~al.} 2022,
  arXiv e-prints, arXiv:2209.09047.
\newblock \doarXiv{2209.09047}

\bibitem[{{Morisset}(2013)}]{morisset13}
{Morisset}, C. 2013, {pyCloudy: Tools to manage astronomical Cloudy
  photoionization code}.
\newblock \doeprint{1304.020}

\bibitem[{{Morrissey} {et~al.}(2018){Morrissey}, {Matuszewski}, {Martin},
  {Neill}, {Epps}, {Fucik}, {Weber}, {Darvish}, {Adkins}, {Allen}, {Bartos},
  {Belicki}, {Cabak}, {Callahan}, {Cowley}, {Crabill}, {Deich}, {Delecroix},
  {Doppman}, {Hilyard}, {James}, {Kaye}, {Kokorowski}, {Kwok}, {Lanclos},
  {Milner}, {Moore}, {O'Sullivan}, {Parihar}, {Park}, {Phillips}, {Rizzi},
  {Rockosi}, {Rodriguez}, {Salaun}, {Seaman}, {Sheikh}, {Weiss}, \&
  {Zarzaca}}]{morrissey18}
{Morrissey}, P., {Matuszewski}, M., {Martin}, D.~C., {et~al.} 2018, \apj, 864,
  93, \dodoi{10.3847/1538-4357/aad597}

\bibitem[{{Murphy} {et~al.}(2021){Murphy}, {Groh}, {Farrell}, {Meynet},
  {Ekstr{\"o}m}, {Tsiatsiou}, {Hackett}, \& {Martinet}}]{murphy21}
{Murphy}, L.~J., {Groh}, J.~H., {Farrell}, E., {et~al.} 2021, \mnras, 506,
  5731, \dodoi{10.1093/mnras/stab2073}

\bibitem[{{Naidu} {et~al.}(2020){Naidu}, {Tacchella}, {Mason}, {Bose}, {Oesch},
  \& {Conroy}}]{naidu20}
{Naidu}, R.~P., {Tacchella}, S., {Mason}, C.~A., {et~al.} 2020, \apj, 892, 109,
  \dodoi{10.3847/1538-4357/ab7cc9}

\bibitem[{{Nanayakkara} {et~al.}(2022){Nanayakkara}, {Glazebrook}, {Jacobs},
  {Bonchi}, {Castellano}, {Fontana}, {Mason}, {Merlin}, {Morishita}, {Paris},
  {Trenti}, {Treu}, {Calabro}, {Boyett}, {Bradac}, {Leethochawalit},
  {Marchesini}, {Santini}, {Strait}, {Vanzella}, {Vulcani}, {Wang}, \&
  {Yang}}]{nanayakkara22}
{Nanayakkara}, T., {Glazebrook}, K., {Jacobs}, C., {et~al.} 2022, arXiv
  e-prints, arXiv:2207.13860.
\newblock \doarXiv{2207.13860}

\bibitem[{{Oey} {et~al.}(2000){Oey}, {Dopita}, {Shields}, \& {Smith}}]{oey00}
{Oey}, M.~S., {Dopita}, M.~A., {Shields}, J.~C., \& {Smith}, R.~C. 2000, \apjs,
  128, 511, \dodoi{10.1086/313396}

\bibitem[{{Offner} {et~al.}(2022){Offner}, {Moe}, {Kratter}, {Sadavoy},
  {Jensen}, \& {Tobin}}]{offner22}
{Offner}, S. S.~R., {Moe}, M., {Kratter}, K.~M., {et~al.} 2022, arXiv e-prints,
  arXiv:2203.10066.
\newblock \doarXiv{2203.10066}

\bibitem[{{O'Sullivan} \& {Chen}(2020)}]{osullivan20}
{O'Sullivan}, D., \& {Chen}, Y. 2020, arXiv e-prints, arXiv:2011.05444.
\newblock \doarXiv{2011.05444}

\bibitem[{{Ouchi} {et~al.}(2009){Ouchi}, {Mobasher}, {Shimasaku}, {Ferguson},
  {Fall}, {Ono}, {Kashikawa}, {Morokuma}, {Nakajima}, {Okamura}, {Dickinson},
  {Giavalisco}, \& {Ohta}}]{ouchi09}
{Ouchi}, M., {Mobasher}, B., {Shimasaku}, K., {et~al.} 2009, \apj, 706, 1136,
  \dodoi{10.1088/0004-637X/706/2/1136}

\bibitem[{{Pahl} {et~al.}(2021){Pahl}, {Shapley}, {Steidel}, {Chen}, \&
  {Reddy}}]{pahl21}
{Pahl}, A.~J., {Shapley}, A., {Steidel}, C.~C., {Chen}, Y., \& {Reddy}, N.~A.
  2021, \mnras, 505, 2447, \dodoi{10.1093/mnras/stab1374}

\bibitem[{{Pauldrach} {et~al.}(2001){Pauldrach}, {Hoffmann}, \&
  {Lennon}}]{pauldrach01}
{Pauldrach}, A.~W.~A., {Hoffmann}, T.~L., \& {Lennon}, M. 2001, \aap, 375, 161,
  \dodoi{10.1051/0004-6361:20010805}

\bibitem[{{Pellegrini} {et~al.}(2011){Pellegrini}, {Baldwin}, \&
  {Ferland}}]{pellegrini11}
{Pellegrini}, E.~W., {Baldwin}, J.~A., \& {Ferland}, G.~J. 2011, \apj, 738, 34,
  \dodoi{10.1088/0004-637X/738/1/34}

\bibitem[{{Pequignot} {et~al.}(1991){Pequignot}, {Petitjean}, \&
  {Boisson}}]{pequignot91}
{Pequignot}, D., {Petitjean}, P., \& {Boisson}, C. 1991, \aap, 251, 680

\bibitem[{{Perez} \& {Granger}(2007)}]{ipython}
{Perez}, F., \& {Granger}, B.~E. 2007, Computing in Science and Engineering, 9,
  21, \dodoi{10.1109/MCSE.2007.53}

\bibitem[{{Ramachandran} {et~al.}(2019){Ramachandran}, {Hamann}, {Oskinova},
  {Gallagher}, {Hainich}, {Shenar}, {Sander}, {Todt}, \&
  {Fulmer}}]{ramachandran19}
{Ramachandran}, V., {Hamann}, W.~R., {Oskinova}, L.~M., {et~al.} 2019, \aap,
  625, A104, \dodoi{10.1051/0004-6361/201935365}

\bibitem[{{Rhode} {et~al.}(2013){Rhode}, {Salzer}, {Haurberg}, {Van Sistine},
  {Young}, {Haynes}, {Giovanelli}, {Cannon}, {Skillman}, {McQuinn}, \&
  {Adams}}]{rhode13}
{Rhode}, K.~L., {Salzer}, J.~J., {Haurberg}, N.~C., {et~al.} 2013, \aj, 145,
  149, \dodoi{10.1088/0004-6256/145/6/149}

\bibitem[{{Robertson} {et~al.}(2015){Robertson}, {Ellis}, {Furlanetto}, \&
  {Dunlop}}]{robertson15}
{Robertson}, B.~E., {Ellis}, R.~S., {Furlanetto}, S.~R., \& {Dunlop}, J.~S.
  2015, \apjl, 802, L19, \dodoi{10.1088/2041-8205/802/2/L19}

\bibitem[{{Sana} {et~al.}(2012){Sana}, {de Mink}, {de Koter}, {Langer},
  {Evans}, {Gieles}, {Gosset}, {Izzard}, {Le Bouquin}, \& {Schneider}}]{sana12}
{Sana}, H., {de Mink}, S.~E., {de Koter}, A., {et~al.} 2012, Science, 337, 444,
  \dodoi{10.1126/science.1223344}

\bibitem[{{Schaerer}(2002)}]{schaerer02}
{Schaerer}, D. 2002, \aap, 382, 28, \dodoi{10.1051/0004-6361:20011619}

\bibitem[{{Schaerer} \& {de Koter}(1997)}]{schaerer97}
{Schaerer}, D., \& {de Koter}, A. 1997, \aap, 322, 598.
\newblock \doarXiv{astro-ph/9611068}

\bibitem[{{Sellmaier} {et~al.}(1996){Sellmaier}, {Yamamoto}, {Pauldrach}, \&
  {Rubin}}]{sellmaier96}
{Sellmaier}, F.~H., {Yamamoto}, T., {Pauldrach}, A.~W.~A., \& {Rubin}, R.~H.
  1996, \aap, 305, L37

\bibitem[{{Sim{\'o}n-D{\'\i}az}(2020)}]{simon-diaz20}
{Sim{\'o}n-D{\'\i}az}, S. 2020, in Reviews in Frontiers of Modern Astrophysics;
  From Space Debris to Cosmology, 155--187, \dodoi{10.1007/978-3-030-38509-5_6}

\bibitem[{{Sim{\'o}n-D{\'\i}az} \& {Stasi{\'n}ska}(2008)}]{simon-diaz08}
{Sim{\'o}n-D{\'\i}az}, S., \& {Stasi{\'n}ska}, G. 2008, \mnras, 389, 1009,
  \dodoi{10.1111/j.1365-2966.2008.13444.x}

\bibitem[{{Skillman} {et~al.}(2013){Skillman}, {Salzer}, {Berg}, {Pogge},
  {Haurberg}, {Cannon}, {Aver}, {Olive}, {Giovanelli}, {Haynes}, {Adams},
  {McQuinn}, \& {Rhode}}]{skillman13}
{Skillman}, E.~D., {Salzer}, J.~J., {Berg}, D.~A., {et~al.} 2013, \aj, 146, 3,
  \dodoi{10.1088/0004-6256/146/1/3}

\bibitem[{{Stark} {et~al.}(2014){Stark}, {Richard}, {Siana}, {Charlot},
  {Freeman}, {Gutkin}, {Wofford}, {Robertson}, {Amanullah}, {Watson}, \&
  {Milvang-Jensen}}]{stark14}
{Stark}, D.~P., {Richard}, J., {Siana}, B., {et~al.} 2014, \mnras, 445, 3200,
  \dodoi{10.1093/mnras/stu1618}

\bibitem[{{Telford} {et~al.}(2021){Telford}, {Chisholm}, {McQuinn}, \&
  {Berg}}]{telford21}
{Telford}, O.~G., {Chisholm}, J., {McQuinn}, K. B.~W., \& {Berg}, D.~A. 2021,
  \apj, 922, 191, \dodoi{10.3847/1538-4357/ac1ce2}

\bibitem[{{The HDF Group}(1997-2020)}]{hdf5}
{The HDF Group}. 1997-2020

\bibitem[{{Tinsley}(1980)}]{tinsley80}
{Tinsley}, B.~M. 1980, \fcp, 5, 287

\bibitem[{{Topping} \& {Shull}(2015)}]{topping15}
{Topping}, M.~W., \& {Shull}, J.~M. 2015, \apj, 800, 97,
  \dodoi{10.1088/0004-637X/800/2/97}

\bibitem[{{Topping} {et~al.}(2022){Topping}, {Stark}, {Endsley}, {Plat},
  {Whitler}, {Chen}, \& {Charlot}}]{topping22}
{Topping}, M.~W., {Stark}, D.~P., {Endsley}, R., {et~al.} 2022, arXiv e-prints,
  arXiv:2208.01610.
\newblock \doarXiv{2208.01610}

\bibitem[{{Vink} {et~al.}(2001){Vink}, {de Koter}, \& {Lamers}}]{vink01}
{Vink}, J.~S., {de Koter}, A., \& {Lamers}, H.~J.~G.~L.~M. 2001, \aap, 369,
  574, \dodoi{10.1051/0004-6361:20010127}

\bibitem[{{Virtanen} {et~al.}(2020){Virtanen}, {Gommers}, {Oliphant},
  {Haberland}, {Reddy}, {Cournapeau}, {Burovski}, {Peterson}, {Weckesser},
  {Bright}, {van der Walt}, {Brett}, {Wilson}, {Millman}, {Mayorov}, {Nelson},
  {Jones}, {Kern}, {Larson}, {Carey}, {Polat}, {Feng}, {Moore}, {VanderPlas},
  {Laxalde}, {Perktold}, {Cimrman}, {Henriksen}, {Quintero}, {Harris},
  {Archibald}, {Ribeiro}, {Pedregosa}, {van Mulbregt}, \& {SciPy 1. 0
  Contributors}}]{scipy2}
{Virtanen}, P., {Gommers}, R., {Oliphant}, T.~E., {et~al.} 2020, Nature
  Methods, 17, 261, \dodoi{10.1038/s41592-019-0686-2}

\bibitem[{{Walborn} {et~al.}(1985){Walborn}, {Nichols-Bohlin}, \&
  {Panek}}]{walborn85}
{Walborn}, N.~R., {Nichols-Bohlin}, J., \& {Panek}, R.~J. 1985, NASA Reference
  Publication, 1155

\bibitem[{{Wang} {et~al.}(2021){Wang}, {Heckman}, {Amor{\'\i}n}, {Borthakur},
  {Chisholm}, {Ferguson}, {Flury}, {Giavalisco}, {Grazian}, {Hayes}, {Henry},
  {Jaskot}, {Ji}, {Makan}, {McCandliss}, {Oey}, {{\"O}stlin}, {Saldana-Lopez},
  {Schaerer}, {Thuan}, {Worseck}, \& {Xu}}]{wang21}
{Wang}, B., {Heckman}, T.~M., {Amor{\'\i}n}, R., {et~al.} 2021, \apj, 916, 3,
  \dodoi{10.3847/1538-4357/ac0434}

\bibitem[{{Wise} \& {Cen}(2009)}]{wise09}
{Wise}, J.~H., \& {Cen}, R. 2009, \apj, 693, 984,
  \dodoi{10.1088/0004-637X/693/1/984}

\bibitem[{{Zastrow} {et~al.}(2013){Zastrow}, {Oey}, \&
  {Pellegrini}}]{zastrow13}
{Zastrow}, J., {Oey}, M.~S., \& {Pellegrini}, E.~W. 2013, \apj, 769, 94,
  \dodoi{10.1088/0004-637X/769/2/94}

\end{thebibliography}
\end{document}